\documentstyle[referee,psfig]{mn}

\footnotesize
\newdimen\digitwidth    %define ! a one digit width for tables
\setbox0=\hbox{\rm0}
\digitwidth=\wd0
\catcode`!=\active
\def!{\kern\digitwidth}
\normalsize

\title[Parkes Multibeam Survey] {The Parkes Multibeam Pulsar Survey: I.
Observing and Data Analysis Systems, Discovery and Timing of 100 Pulsars}
\author[R. N. Manchester et al.]
{R. N. Manchester,$^1$\thanks{Email: rmanches@atnf.csiro.au}
A. G. Lyne,$^2$
F. Camilo,$^{2,3}$
J. F. Bell,$^1$
V. M. Kaspi,$^{4,5}$
\newauthor
N. D'Amico,$^{6,7}$
N. P. F. McKay,$^2$
F. Crawford,$^5$
I. H. Stairs,$^{2,8}$
\newauthor
A. Possenti,$^6$
M. Kramer,$^2$
and D. C. Sheppard$^2$
\\
$^1$ Australia Telescope National Facility, CSIRO, P.O.~Box~76, Epping
NSW~1710, Australia\\
$^2$ University of Manchester,
Jodrell Bank Observatory, Macclesfield, Cheshire, SK11~9DL, UK\\
$^3$ Columbia Astrophysics Laboratory, Columbia University, 550 W.
120th Street, New York, NY 10027, USA\\
$^4$ McGill University, Ernest Rutherford Physics Building, 3600 University
Street, Montreal, QC, Canada H3A 2T8  \\
$^5$ Massachusetts Institute of Technology, Center
for Space Research, 70 Vassar Street, Cambridge, MA~02139, USA \\
$^6$ Osservatorio Astronomico di Bologna, via Ranzani 1, 40127
Bologna, Italy\\
$^7$ Istituto di Radioastronomia del CNR, via Gobetti 101, 40129
Bologna, Italy\\
$^8$ National Radio Astronomy Observatory, Green Bank, WV 24944, USA
}

\date{Received by  MNRAS on December 11, 2000. Revised version accepted on June 14, 2001}
\begin{document}

\maketitle
\newcommand{\setthebls}{
%                 de-comment this line for double spacing:
%\baselineskip=20pt
}

\setthebls

\begin{abstract} 
The Parkes multibeam pulsar survey is a sensitive survey of a strip along
the Galactic plane with $|b|<5\degr$ and $l=260\degr$ to $l=50\degr$. It
uses a 13-beam receiver on the 64-m Parkes radio telescope, receiving two
polarisations per beam over a 288 MHz bandwidth centred on 1374
MHz. Receiver and data acquisition systems are described in some detail. For
pulsar periods in the range 0.1 -- 2 s and dispersion measures of less than
300 cm$^{-3}$ pc, the nominal limiting flux density of the survey is about
0.2 mJy. At shorter or longer periods or higher dispersions, the sensitivity
is reduced. Timing observations are carried out for pulsars discovered in
the survey for 12 -- 18 months after confirmation to obtain accurate
positions, spin parameters, dispersion measures, pulse shapes and mean flux
densities. The survey is proving to be extremely successful, with more than
600 pulsars discovered so far. We expect that, when complete, this one
survey will come close to finding as many pulsars as all previous pulsar
surveys put together. The newly discovered pulsars tend to be young, distant
and of high radio luminosity. They will form a valuable sample for studies
of pulsar emission properties, the Galactic distribution and evolution of
pulsars, and as probes of interstellar medium properties. This paper reports
the timing and pulse shape parameters for the first 100 pulsars timed at
Parkes, including three pulsars with periods of less than 100 ms which are
members of binary systems. These results are briefly compared with the
parameters of the previously known population.
\end{abstract}


\section{INTRODUCTION}\label{sec:intro}
Since the discovery of pulsars more than 30 years ago \cite{hbp+68}, many
different searches for these objects have contributed to the 730 or so
pulsars known prior to mid-1997 when the survey described here commenced. Some efforts with a
relatively narrow focus have resulted in the discovery of extremely
important objects, for example, the Crab pulsar \cite{sr68} or the first
millisecond pulsar \cite{bkh+82}. However, the vast majority of known
pulsars have been found in larger-scale searches. These searches generally
have well-defined selection criteria and hence provide samples of the
Galactic population which can be modeled to determine the properties of the
parent population. Most of our knowledge about the Galactic distribution and
the evolution of pulsars has come from such studies (e.g. Lyne, Manchester
\& Taylor 1985, Lorimer et al. 1993, Hartman et al. 1997, Cordes \& Chernoff
1998, Lyne et al. 1998)\nocite{lmt85,lbdh93,hbwv97a,cc98,lml+98}. Of
particular significance are young pulsars. These are often associated with
supernova remnants (e.g. Kaspi 2000)\nocite{kas00}, show significant period
irregularities such as glitches \cite{lsg00} and have pulsed emission
at optical, X-ray and $\gamma$-ray wavelengths (e.g. Wallace et al. 1977,
Thompson et al. 1999).\nocite{wpm+77,tbb+99}
  
Of comparable importance though, are the serendipitous discovery of
unusual and often unique objects by larger-scale surveys. Examples of
this abound --- for example, the first binary pulsar, PSR B1913+16
\cite{ht74}, the first star with planetary-mass companions
\cite{wf92}, the first pulsar with a massive stellar companion
\cite{jml+92}, and the first eclipsing pulsar \cite{fst88}. Pulsars
show an amazingly diverse range of properties and most major surveys
turn up at least one object with new and unexpected
characteristics. Some of these are of great significance.  The prime
example is of course PSR B1913+16, which has provided the first
observational evidence for gravitational waves and the best evidence
so far that general relativity is an accurate description of
gravity in the strong-field regime \cite{tw89}.

Pulsars are relatively weak radio sources. Successful pulsar surveys
therefore require a large radio telescope, low-noise receivers, a
relatively wide bandwidth and long observation times. Pulsar signals suffer
dispersion due to the presence of charged particles in the interstellar medium. The dispersion
delay across a bandwidth of $\Delta\nu$ centred at a frequency $\nu$ is
\begin{equation}\label{eq:dm}
\tau_{\rm DM} = 8.30 \times 10^3\,{\rm DM}\,\Delta\nu\,\nu^{-3}\;\;{\rm s},
\end{equation}
where the dispersion measure, DM, is in units of cm$^{-3}$ pc and the
frequencies are in MHz. To retain sensitivity, especially for short-period,
high-dispersion pulsars, the observing bandwidth must be sub-divided into
many channels. In most pulsar searches to date, this has been achieved using
a filterbank system.

The sensitivity of pulsar searches is also limited by the Galactic radio
continuum background and by interstellar scattering, especially for low
radio frequencies and at low Galactic latitudes. Interstellar scattering
results in a one-sided broadening of the observed pulse profile with a frequency
dependence $\sim \nu^{-4.4}$ (e.g. Rickett 1977)\nocite{ric77} which cannot
be removed by using narrow bandwidths. Most pulsar searches along the
Galactic plane have therefore been at higher radio frequencies, often around
1400 MHz (e.g. Clifton et al. 1992, Johnston et
al. 1992).\nocite{clj+92,jlm+92}

The Clifton et al. 1400 MHz survey was carried out using the 76-m Lovell Telescope at
Jodrell Bank Observatory, and covered a strip along the Galactic plane with
$|b| < 1.1\degr$ between longitudes of $355\degr$ and 95\degr, with a narrower
extension to 105\degr. The limiting sensitivity to long-period pulsars away
from the Galactic plane was about 1 mJy. A total of 61 pulsars was
detected, of which 40 were not previously known. Johnston et al. carried out
a complementary survey of the southern Galactic plane in the region $|b| <
4\degr$ and between $l=270\degr$ and $l=20\degr$, with a central frequency
of 1500 MHz. The limiting sensitivity was very similar to that for the
Clifton et al. survey. A total of 100 pulsars was detected of which 46 were
previously unknown. These surveys found a sample of young and generally
distant pulsars which are strongly concentrated at low Galactic longitudes,
$|l|\la40\degr$. They include a number of interesting objects, including the
eclipsing high-mass binary system PSR B1259$-$63 \cite{jml+92} and many
glitching pulsars \cite{sl96,wmp+00}. 

The Parkes multibeam receiver was conceived with the aim of undertaking
large-scale and sensitive searches for relatively nearby galaxies ($z \la
0.04$) by detection of their emission in the 21-cm line of neutral
hydrogen. The receiver has 13 feeds with a central feed surrounded by two
rings, each of six feeds, arranged in a hexagonal pattern
\cite{swb+96}. This arrangement permits the simultaneous observation of 13
regions of sky, increasing the speed of surveys by approximately the same
factor. It was quickly realised that this system would make a powerful
instrument for pulsar surveys, provided the bandwidth was increased above
the original specification and the necessary large filterbank system could
be constructed. A new data acquisition system capable of handling multibeam
data sets was also a fundamental component of the system.

These requirements were met, and the Parkes multibeam pulsar survey
commenced in August 1997. This survey aims to cover a strip with
$|b|<5\degr$ along the Galactic plane between Galactic longitudes of
260\degr~and 50\degr. The filterbank system gives $96 \times 3$ MHz channels
of polarisation-summed data for each beam which are sampled every 250
$\mu$s. Observation times per pointing are 35 min, giving a very high
sensitivity, about seven times better than those of the Clifton et
al. (1992) and Johnston et al.  (1992) surveys, at least for pulsars not in
short-period binary systems. Although not yet complete, the survey has been
outstandingly successful, with over 600 pulsars discovered so far.

Preliminary reports on the multibeam survey and its results have been given
by Camilo et al. (2000a),\nocite{clm+00} Manchester et
al. (2000),\nocite{mlc+00} Lyne et al. (2000)\nocite{lcm+00} and D'Amico et
al. (2000)\nocite{dlm+00}. Also, papers on the discovery of several pulsars
of particular interest have been published. Lyne et al. (2000) announced the
discovery of PSR J1811$-$1736, a pulsar with a period of 104 ms in a highly
eccentric orbit of period 18.8 d with a companion of minimum mass 0.7
M$_{\odot}$, most probably a neutron star, making this the fourth or fifth
double-neutron-star system known in the Galactic disk. Camilo et
al. (2000b)\nocite{ckl+00} report the discovery of two young pulsars,
J1119$-$6127 and J1814$-$1744, which have the highest surface dipole
magnetic field strengths among known radio pulsars. PSR J1119$-$6127 has a
characteristic age, $\tau_c$, of only 1600 years, a measured braking index,
$n = 2.91 \pm 0.05$ and is associated with a previously unknown supernova
remnant \cite{cgk+01,pkc+01}. PSR J1814$-$1744 has a much longer period,
3.975~s, and the highest inferred surface dipole field strength of any known
radio pulsar, $5.5 \times 10^{13}$~G, in the region of so-called
``magnetars'' \cite{pkc00}. PSR J1141$-$6545 is a relatively young pulsar
($\tau_c \sim 1.4$ Myr) in an eccentric 5-hour orbit for which the
relativistic precession of periastron has been measured \cite{klm+00a}. This
implies that the total mass of the system is 2.30 M$_{\odot}$, indicating
that the companion is probably a massive white dwarf formed before the
neutron star we observe as the pulsar. Stairs et al. (2001)\nocite{sml+01}
discuss the high-mass binary system PSR J1740$-$3052 which is in a highly
eccentric 230-day orbit with a companion star of minimum mass 11
M$_{\odot}$. A possible companion is a late-type star identified on infrared
images, but the absence of the expected eclipses and precession of
periastron due to tidal interactions suggest that the actual companion may
be a main-sequence B-star or a black hole hidden by the late-type
star. Camilo et al. (2001)\nocite{clm+01} report the discovery of five
circular-orbit binary systems with orbital periods in the range 1.3 -- 15
days. Three of these pulsars, PSRs J1232$-$6501, J1435$-$6100 and
J1454$-$5846, as well as PSR J1119$-$6127, were discovered early in the
survey and hence are included in the pulsars described in this
paper. Finally, D'Amico et al. (2001)\nocite{dkm+01} report the discovery of
two young pulsars, PSRs J1420$-$6048 and J1837$-$0604, which may be
associated with EGRET $\gamma$-ray sources.

In the following section we describe the observing and analysis systems and
the search strategy. Timing observations undertaken after the confirmation
of a pulsar and our data release policy are described in Section 3. In
Section 4, we give parameters for the first 100 pulsars discovered by the
survey.
Implications of these results are discussed in Section 5. Detailed
information about the survey, observing instructions, data release policy,
and results may be found under the pulsar multibeam web page.\footnote{
http://www.atnf.csiro.au/research/pulsar/pmsurv/.} 

\section{OBSERVING AND SEARCH ANALYSIS SYSTEMS}
In this section, we describe in detail the receiver system, data acquisition
system, analysis procedures and search strategy being used for the Parkes
multibeam pulsar survey.

\subsection{The Receiver System}\label{sec:rcvr}
The Parkes multibeam receiver consists of a 13-feed system operating at a
central frequency of 1374 MHz with a bandwidth of 288 MHz at the prime focus
of the Parkes 64-m radio telescope. Orthogonal linear polarisations are
received from each feed and fed to cryogenically cooled HEMT amplifiers,
constructed under contract at Jodrell Bank Observatory. The horns are
arranged in a double hexagon around a central horn with a spacing between
horns of 1.2 wavelengths; the corresponding beam spacing on the sky is close
to twice the nominal half-power beamwidth of 14.2 arcmin
\cite{swb+96}. Measured system parameters\footnote{From
http://www.atnf.csiro.au/research/multibeam/lstavele/description.html.} are
listed in Table~\ref{tb:rcvr}.  System temperatures vary by a degree or so
over the 26 receivers; the value of 21 K quoted in the table is an average
value. For the central beam, this corresponds to an equivalent system flux density
of 28.6 Jy. Outer feeds have a somewhat lower efficiency, reduced by
about 0.27 db for the inner ring and 1.0 db for the outer ring. The outer
beams are also somewhat elliptical, with the major axis in the radial
direction, and have a significant coma lobe. Predicted beam patterns for the
central and outer beams are given by Staveley-Smith et al. (1996); at least
to the half-power point, the beam patterns are well represented by a two-dimensional
Gaussian function.

\begin{table}
\caption{Feed and receiver parameters}
\begin{tabular}{llll}
\hline
Number of beams & 13 & & \\
Polarisations/beam & 2 & & \\
Frequency channels/polarisation & $96\times 3$ MHz \\
System temperature (K)  & 21 & &  \\ \\
Beam & Centre & Inner Ring & Outer Ring \\
Telescope gain (K/Jy) & 0.735 & 0.690 & 0.581 \\
Half-power beamwidth (arcmin) & 14.0 & 14.1 & 14.5 \\
Beam ellipticity & 0.0 & 0.03 & 0.06 \\
Coma lobe (db) & none & $-17$ & $-14$ \\
\hline
\end{tabular}
\label{tb:rcvr}
\end{table}

After further amplification, all 26 signals are down-converted in the focus
cabin to intermediate frequency using a
local oscillator frequency of 1582 MHz. These signals are transferred to the
tower receiver room via low-loss coaxial cables and pass through
cable-equalising amplifiers and level setting attenuators to a
down-conversion system. This splits the 288-MHz bandwidth of each signal
into three equal parts with output between 64 and 160 MHz using an up-down
conversion system with band-limiting filters centred at 1060 MHz. These
signals are then fed to a very large filterbank system, designed and
constructed at Jodrell Bank Observatory and Osservatorio Astronomico di
Bologna, which gives 96 3-MHz channels for each polarisation of each
feed. The output of each filter is detected and summed with its
corresponding polarisation pair. These summed outputs are high-pass filtered
with an effective time constant of approximately 0.9 s, integrated for the
sampling interval of 250 $\mu$s and then one-bit digitised.

\subsection{Data Acquisition and Analysis}\label{sec:data}
Data acquisition is controlled by a multi-threaded C++ program, {\sc pmdaq},
running on a Digital Alpha {\sc picmg} processor.  A custom-designed board
with a programmable Xilinx device is installed on the computer's PCI bus,
and interfaces between the digitiser and an Ikon-10116 16-bit direct memory
access card. Integration of the first sample of an observation is triggered
by the Observatory 1-s pulse, allowing measurement of pulse arrival
times. The first 16-bit word of every input sample is a counter which is
checked by the data acquisition program and then discarded. Time
synchronisation is further checked by using a 5-s pulse from the Observatory
clock. Data can be output to disk, double-density Exabytes or digital linear
tapes (DLTs). Each output block contains a 640-byte header giving telescope,
receiver, source and observation parameters and 48 kbyte of one-bit data,
all from a single beam. Successive blocks have data from successive
beams. Survey-mode data are normally output to DLTs and timing data to
Exabytes. For survey observations, the data rate is 640 kbyte s$^{-1}$,
which fills a DLT in approximately 15 hours of continuous observation.

Observations are controlled using a Tcl-Tk interface to a control program,
{\sc pmctrl}, operating on a Sun Sparc workstation. The interface allows
setting of observation parameters such as the receiver, filterbank system,
sampling interval, observation time, output device, pointing centre and feed
position angle and the logging of operator messages. {\sc pmctrl} has socket
interfaces to the Observatory clock, the telescope drive system and the
receiver translator system and an RPC interface to PMDAQ.
The program maintains a record of tape operations and
handles status returns and error conditions from the telescope or data
acquisition system. It also writes a summary observation file and a complete
log file giving details of all observations.  Details of the observing
strategy for the multibeam survey are given in \S\ref{sec:strategy}.

Observations can be monitored in real time using a program, {\sc pmmon},
which runs on a networked workstation with user input via a Tcl-Tk
interface. {\sc pmmon} communicates with {\sc pmdaq} via an RPC interface,
obtaining either complete tape blocks or data streams summed across all
filter channels for each beam. Several forms of output are provided,
including mean digitiser levels for each beam, modulation spectra and time
sequences for each beam, and modulation spectra for each filterbank channel
of a given beam. The latter form of output is especially valuable for
tracing narrow-band interference. For observations of known pulsars
(normally with the centre beam), integrated pulse profiles for each
frequency channel and a dedispersed mean pulse profile can be displayed and
may be recorded to disk for later examination.

Offline processing runs on networked workstations at each of the collaborating institutions under the control of a
Java program, {\sc pmproc}. The processing consists of four main stages. Data are first examined for
the presence of narrow-band radio-frequency interference by computing the
modulation spectrum for each frequency channel, normally using a subset of
each data file of length $2^{19}$ samples. Samples in channels
containing strong interference are set to zero or one in alternate channels
(to give a mean of 0.5) as the data are transferred to disk in subsequent
stages.

The second stage of processing concerns identification of interfering
signals in the modulation spectrum. Since most interference is undispersed,
this analysis is performed on the `zero-DM' spectrum. Data for each
observation are summed across all frequency channels on reading from the
tape to produce a zero-DM data stream of $2^{23}$ samples per beam. This is
Fourier-transformed to give the modulation spectrum. Known signals which are
present all or most of the time, such as the power line frequency (50 Hz)
and its harmonics, are first identified and their bandwidth determined. The
remaining spectrum is then searched for significant spectral features. This
search is performed on the fundamental spectrum and on spectra obtained by
summing 2, 4, 8 and 16 harmonics. Signals are identified and their bandwidth
and harmonic content recorded. Any signal which appears in four or more
beams of a given pointing is flagged as interference; that signal and its
harmonics are deleted in subsequent processing steps for that
pointing. Similarly, any signal which appears in a given beam in more than
three pointings is marked for deletion in subsequent processing for that
beam in all pointings on that tape, and any signal which appears more than
seven times in any beam of a given tape is marked for deletion in all
pointings on that tape. A summary output is produced for each tape (normally
containing 20 -- 25 pointings) which gives grey-scale images of the
modulation spectra as a function of beam and pointing, and lists the
frequency ranges identified as interference.

In the third and major stage of processing, the data are searched for
periodic signals over a range of dispersion delays. The basic analysis
procedure is very similar to that employed in the Parkes Southern pulsar
survey and described in detail by Manchester et al. (1996).\nocite{mld+96} A
`tree' dedispersion algorithm \cite{tay74} is used. Dispersion delays are
proportional to $\nu^{-2}$, but the tree algorithm assumes that they are
linear with frequency. This is approximately true for small fractional
bandwidths, but the multibeam survey has a fractional bandwidth of
about 20 per cent, and straightforward application of tree dedispersion
would lead to excessive pulse smearing for short-period pulsars. Also, the
tree algorithm requires a number of frequency channels which is a power of
two. To overcome these problems, the delays are `linearised' on reading from
tape. The number of frequency channels is increased from 96 to 128, and
channel data streams are reassigned in channel number to remove the
second-order dispersion-delay term. These channel reassignments are
independent of dispersion measure.

The linearised data are split into 8 sub-bands, each of 16 channels. A tree
dedispersion is performed on each of these sub-bands to give dedispersed
data streams for 16 dispersions between zero and the `diagonal DM' (at which
the dispersion smearing across one channel equals the sampling interval),
approximately 35 cm$^{-3}$ pc.  These are subsequently added with varying
delays to give a range of DMs about the central value. Another application
of the tree algorithm to delayed data gives a further 16 data streams for
dispersions from 35 to 70 cm$^{-3}$ pc. Data samples are then summed in
pairs to give an effective sampling interval of 0.5 ms and the tree
algorithm is applied again to give 16 data streams for dispersions from 70
to 139 cm$^{-3}$ pc. This process is repeated up to four more times, to an
effective sampling interval of 8 ms, until a maximum DM of 2177 cm$^{-3}$ pc
or 42/$\sin |b|$ cm$^{-3}$ pc, where $b$ is the Galactic latitude, whichever
is less, is reached. The dedispersed data streams for each sub-band are then
summed with a range of delays to give up to 325 dedispersed data streams
with DM in the range 0 to 2203 cm$^{-3}$ pc. The DM steps are 0.54 cm$^{-3}$
pc for the first tree data set, 0.81 cm$^{-3}$ pc for the second,
and 26 cm$^{-3}$ pc for the last,
increasing by roughly a factor of two for each successive tree data set
after the second.

For each DM, the summed data stream is high-pass filtered by subtracting a
running mean of length 2.048 s and then Fourier-transformed using a fast
Fourier transform (FFT) routine. After deletion of spectral channels
affected by interference and interpolation to recover spectral features
lying midway between Fourier bins, the resulting spectra are searched for
significant peaks. This process is repeated for spectra in which 2, 4, 8 and
16 harmonics have been summed to give a set of 50 candidate periods (10 from
the fundamental and from each harmonic sum) for each DM.  A pulse profile is then
formed for each candidate period by inverse transformation of the complex
Fourier components for the fundamental and its harmonics, and the
signal-to-noise ratio of this profile computed. All such profiles from the
full analysis over all DMs for a given beam are then ordered by
signal-to-noise ratio. For the top 66 candidates, the appropriate tree data
streams are summed into 4 sub-bands and folded into 16 sub-integrations,
each of duration a little over 2 min, using the nominal period and DM. These
are then summed with a range of delays in frequency and time, up to one
sample per sub-band and per sub-integration respectively, to search for the
highest signal-to-noise ratio over a range of period and DM about the
nominal values. The candidate parameters, including the maximum
signal-to-noise ratios obtained from the harmonic summing, the reconstructed
profile and results from the $P$--DM search are then recorded for later examination.

\begin{figure}
\centerline{\psfig{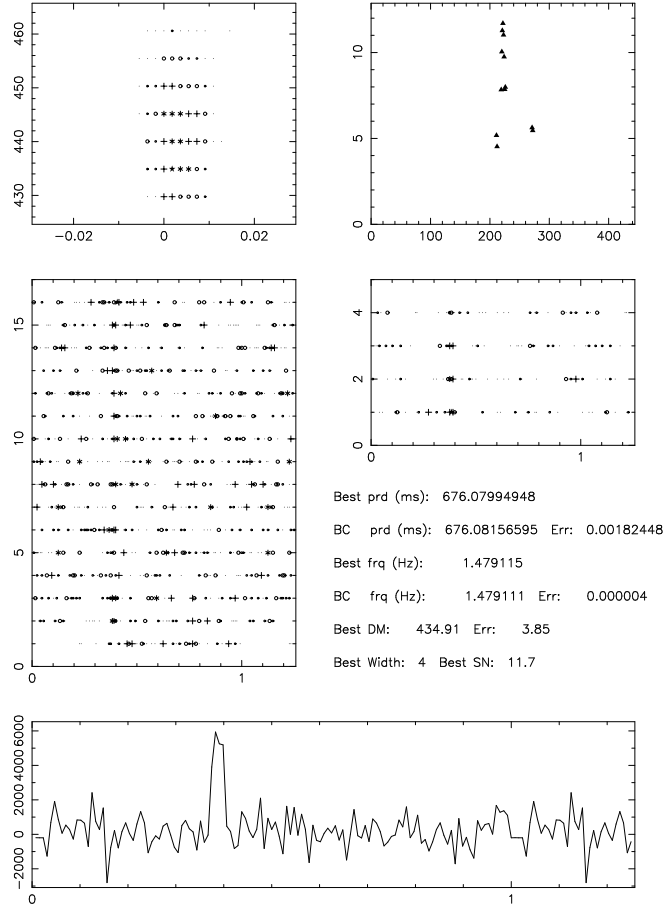}} 
\caption{Display plot for typical candidate, later confirmed as a pulsar,
PSR J1056$-$5709. Clockwise from the top left, the sub-plots show a crude
greyscale of the dependence of signal-to-noise ratio on dedispersion DM and
offset (in ms) from the nominal period, the dependence of signal-to-noise
ratio on DM trial number, a greyscale plot of signal-to-noise ratio versus pulse phase
for 4 sub-bands across the observed bandwidth of 288 MHz, the final mean
pulse profile, and a greyscale plot of signal-to-noise ratio versus pulse
phase for successive sub-integrations, each of approximately 2 min duration.}
\label{fg:cand}
\end{figure}

In the next stage of processing, candidates from all pointings on a given
tape are collated and searched for common periods. Candidate periods seen in
more than 6 beams are rejected as interference. Remaining candidates with a
$P$--DM signal-to-noise ratio above a threshold (normally 8.0, corresponding
to a random occurrence every few beams) are then examined using an
interactive display and classified as Class 1 or Class 2 candidates or
rejected as probable interference. Fig.~\ref{fg:cand} shows the display plot
for a typical Class 1 candidate, later confirmed as a pulsar. The
classification is necessarily somewhat subjective and is based on the
similarity of the subplots to those for known pulsars. The most important
criteria are final signal-to-noise ratio, continuity across sub-integrations
and sub-bands of the pulse signal, and a well-defined peak in
signal-to-noise ratio versus DM. The signal should also be linear or
parabolic (indicating a constant acceleration) in the phase-time plot and
linear in the phase-frequency plot. Most Class 1 candidates have a
signal-to-noise ratio of 10 or more. For the early low-latitude phases of
the survey, a Class 1 candidate was selected every one or two
pointings. Each candidate is identified by a unique code based on the
processing centre and a sequential number.

Candidates are then re-observed using the centre beam of the multibeam
receiver in order to confirm their reality as pulsars. Observations are made
at five grid positions, the nominal position and four positions offset in
latitude and longitude by 9 arcmin, normally with 6 min integration per
point. These observations are searched in period and DM about the nominal
values and, if two or three detections are obtained, an improved position is
computed from the relative signal-to-noise ratio. If there is no detection
in the grid observations, a 35-min observation is made at the nominal
position and searched for a significant signal. This search is usually made
using Fourier techniques to detect pulsars whose period may have changed
significantly from the nominal value, due to binary motion for
example. Candidates which are not redetected in one or two such observations
are down-graded or rejected. To date, all Class 1 candidates have been 
re-observed with about 80 per cent of them being confirmed as pulsars. 

\begin{table}
\caption{Pulsar multibeam survey parameters}
\begin{tabular}{ll}
\hline
Galactic longitude range & 260\degr~ to 50\degr \\
Galactic latitude range  & $-5$\degr~ to 5\degr \\
Hexagonal grid spacing 	  & $0\fdg2333$  \\
Number of survey pointings & 2670 \\
Sampling interval, $\tau_{samp}$ & 250 $\mu$s \\
Observation time/pointing, $\tau_{obs}$ & 2100 s \\
Limiting sensitivity for centre beam & 0.14 mJy \\ 
\hline
\end{tabular}
\label{tb:survey}
\end{table}

\subsection{Survey Sensitivity}
\label{sec:sens}
Survey parameters are summarised in Table~\ref{tb:survey}. The system
sensitivity for the centre beam has been modeled by Crawford
(2000)\nocite{cra00}, assuming the parameters given in Tables \ref{tb:rcvr}
and \ref{tb:survey}. The raw limiting flux density is given by the
radiometer equation
\begin{equation}
S_{lim} = \frac{\sigma \beta T_{sys}}{G \sqrt{B N_p \tau_{obs}}}
\end{equation}
where $\sigma$ is a loss factor, taken to be 1.5,\footnote{One-bit sampling
at the Nyquist rate introduces a loss of $\sqrt{2/\pi}$ relative to a fully
sampled signal (cf. Van Vleck \& Middleton 1966)\nocite{vm66}. The principal
remaining loss results from the non-rectangular bandpass of the channel
filters.}  $\beta$ is the detection signal-to-noise ratio threshold, taken
to be 8.0, $T_{sys}$ is the system temperature, $G$ is the telescope gain,
$B$ is the receiver bandwidth in Hz, $N_p$ is the number of polarisations
and $\tau_{obs}$ is the time per observation in seconds.

An idealised pulse train of frequency $f_1=P^{-1}$, where $P$ is the pulse
period, is represented in the Fourier domain by its fundamental and 15
harmonics $F(f_i)$, where each of the harmonics has an amplitude $y_0(f_i)=
1/S_{lim}$. These harmonics are then multiplied by a series of functions,
representing the responses of the various filters in the system, to give a
final set of Fourier amplitudes $y(f_i)$. The first filter function is the
Fourier transform of the intrinsic pulse profile, assumed to be Gaussian
with a half-power width of $W_{50}=0.05 P$,
\begin{equation}
|g_1(f)| = \exp\left(-\frac{\pi^2 f^2 W_{50}^2}{4 \ln 2}\right)
\end{equation}
and by a similar function $g_2(f)$ representing the Fourier transform of the
smearing due to dispersion in each filter channel, also assumed to
have a Gaussian response, with $W_{50}$ replaced by $\tau_{\rm DM}$
(Equation~\ref{eq:dm}). Since the analysis is based on
the amplitude spectrum and each of the filters is real, we only have to consider the
amplitude response of each filter. 

The harmonics are then multiplied by the Fourier
response of each of the filters in the hardware and software system. These
result from the finite sampling interval,
\begin{equation}
|g_3(f)| = \left|\frac{\sin(\pi f \tau_{samp})}{\pi f \tau_{samp}}\right|,
\end{equation}
the digitiser high-pass filtering, a two-pole filter with amplitude response
\begin{equation}
|g_4(f)| = \frac{(2\pi f \tau_{\rm HP})^2}{[1+(2\pi f \tau_{\rm HP})^4]^{1/2}},
\end{equation}
where $\tau_{\rm HP} = 0.9$ s (see \S\ref{sec:100psrs}), and a software
high-pass filter, implemented by subtracting a box-car average of length
$\tau_S = 2.048$ s from the dedispersed data stream, giving
\begin{equation}
|g_5(f)| = 1 - \frac{\sin(\pi f \tau_S)}{\pi f \tau_S}.
\end{equation}

The harmonic range is then limited to $f > f_{min}$, where $f_{min} = 0.2$
Hz, a limit set mainly by the need to reject low-level interference and
other red noise, and $f < f_N = 1/(2 \tau_{samp})$, the Nyquist
frequency. Harmonics of the lowest valid signal frequency are then summed to
give a final amplitude
\begin{equation}
Y(f_n) = \frac{\sum_{i=1}^{n} y(f_i)}{\sqrt{n}}.
\end{equation}
The final limiting sensitivity $S_{min}$ is then given by
\begin{equation}
S_{min} = \frac{1}{Y_{max}(f_n)},
\end{equation}
where $Y_{max}(f_n)$ is the largest $Y(f_n)$ for $n$ = 1, 2, 4, 8 or 16.

\begin{figure} 
\centerline{\psfig{file=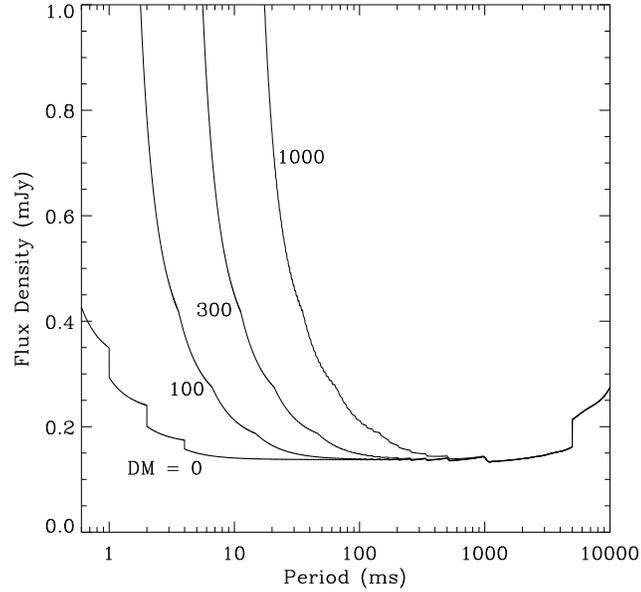,width=90mm}} 
\caption{Minimum detectable flux density for the Parkes multibeam survey as
a function of pulsar period and DM. These calculations refer to the centre
of the central beam of the multibeam system, are for an assumed pulse
width of $0.05 P$ and do not include the effects of increased system
temperature due to the Galactic sky background emission or the effects of
interstellar scattering or interference. }
\label{fg:sens}
\end{figure}

The resultant sensitivity curves for four representative values of DM are
shown in Fig.~\ref{fg:sens}. These curves show that for low-DM pulsars with
periods greater than about 10 ms, the limiting sensitivity is about 0.14
mJy. Steps in the zero-DM curve at short periods result from changes in the
number of harmonics below the Nyquist frequency; at higher DMs, the higher
harmonics are attenuated and the steps are not as evident. Steps between 100
ms and 1 s result from the software high-pass filtering. The Fourier
cutoff at $f_{min}$ and the hardware and software high-pass filtering
results in reduced sensitivity at longer periods. 

Especially for distant pulsars near the Galactic plane, the sensitivity is
degraded by two effects not included in the modeling: sky background
temperature ($T_{sky}$) and pulse smearing due to scattering
($\tau_{scatt}$). Limiting sensitivities should be scaled by factors
$(T_{sys} + T_{sky})/T_{sys}$ and $[w/(P-w)]^{1/2}/[w_0/(P-w_0)]^{1/2}$,
where $w = (W_{50}^2 + \tau_{samp}^2 + \tau_{\rm DM}^2 +
\tau_{scatt}^2)^{1/2}$ is the effective pulse width, $W_{50}$ is the
intrinsic pulse width, and $w_0 = [(0.05P)^2 +
\tau_{samp}^2 + \tau_{\rm DM}^2]^{1/2}$. Sky background temperatures are
highest close to the Galactic plane and towards the Galactic Centre; for
example at ($l,b = 300\degr, 0\degr$), $T_{sky} \sim 5$ K and for ($l,b =
350\degr, 0\degr$), $T_{sky} \sim 18$ K. Scattering parameters have not yet
been measured for the multibeam pulsars, but a cursory examination of the
mean pulse profiles shows that at least 15 per cent have scattering
broadening of a few milliseconds or more.

It should also be emphasised that these sensitivity figures refer to centre
of the central beam. As Table~\ref{tb:rcvr} shows, the outer beams are less
sensitive. Averaged over the 13 beams, the limiting sensitivity is about
0.16 mJy. Also, of course, pulsars do not usually lie at the beam centre in
the discovery observation. The limiting sensitivity is further degraded by
the beam response at the position of the pulsar relative to that at the beam
centre. The average beam gain over the hexagonal area covered by one beam
(see Section~\ref{sec:strategy} below) assuming a gaussian beamshape,
is 0.70, giving an average limiting
flux density for the survey as a whole of 0.22 mJy.

The sensitivity is also degraded by radio frequency interference, but this
is much more difficult to quantify. There are many forms of interference,
including both natural and man-made signals. Natural interference such as
lightning is not a major problem as it is not periodic and some protection
is afforded by the one-bit digitisation. Some of the man-made interference
originates from within the Observatory and even from within the receiving
system itself, but most sources are narrow-band transmissions such as radar
beacons and communication links. Much of the interference is transient,
which makes it difficult to trace. Typically 6 -- 8 frequency channels are
routinely rejected because they contain persistent modulated narrow-band
signals. The sensitivity of the system to modulation at the power-line
frequency (50 Hz) was minimised by choosing a sampling interval such that
the Nyquist frequency is a harmonic of 50 Hz. Although not strictly
interference, beam 8A has been disconnected since the start of the survey
because of a quasi-periodic gain modulation occurring in the cryogenically
cooled part of the receiver.  Also, coupling within the one-bit digitiser
results in periodic signals at frequencies of $f_N/2^n$, where $n$ is an
integer, and their harmonics. These are rejected in the Fourier
domain. After rejection of the known sources of interference, typically
there are 20 -- 30 narrow-band signals (`birdies') detected in the zero-DM
modulation spectra for a full tape. These are flagged and deleted from the
pointings in which they were detected. Typically, much less than one per
cent of the modulation spectrum is rejected.

\subsection{Search Strategy}\label{sec:strategy}
\begin{figure} 
\centerline{\psfig{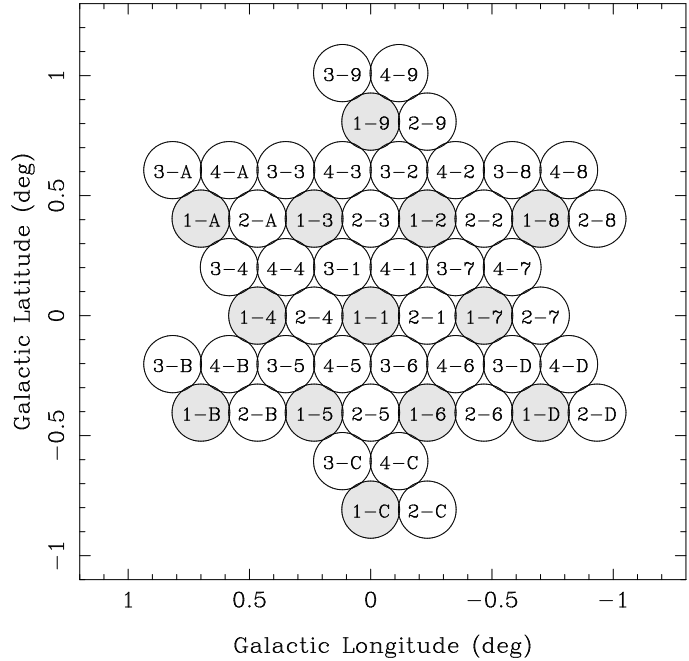}} 
\caption{Beam locations for a cluster of four pointings for a feed
Galactic position angle of 30\degr. Beams are labeled with a pointing
number within the cluster and the hexadecimal beam number; beams for
pointing 1 are shaded. }
\label{fg:beams}
\end{figure}

The 13 beams of the multibeam receiver are spaced by approximately two
beamwidths on the sky. Therefore interleaved pointings are required to cover
a given region. As shown in Fig.~\ref{fg:beams}, a cluster of four pointings
covers a region about 1.5\degr~ across with adjacent beams touching at the
half-power points. Clusters tessellate to fully cover a region. For this
configuration, the multibeam receiver must be oriented at a Galactic
position angle of 30\degr. Since the time per pointing is relatively long
(35 min), the variation of parallactic angle is tracked during the
observation. The range of parallactic angle is $\pm 180\degr$ but the
multibeam receiver has a feed-angle range limited to $\pm 75\degr$, and so
$\pm 60\degr$ or $\pm 120\degr$ may be added to the feed angle to keep it
within the legal range throughout the observation. This changes the labels
on the beams in Fig.~\ref{fg:beams} but not the pattern.

The survey region, $-100\degr < l < 50\degr$ and $|b| < 5\degr$, is covered by
a grid of survey pointings, defined by
\begin{eqnarray}
l & = & (i_l-5000+0.5\,i_{b2})\,d_l~{\rm and} \\
b & = & (i_b-500)\,d_b,
\end{eqnarray}
where
\begin{eqnarray}
i_l & = & 4400+7n+2m+c_l, \\
i_b & = & 500-2n-8m+c_b,
\end{eqnarray}
$d_l = 0.5\,\Delta$, $d_b = 0.5\,\Delta\,${\rm sin}\,60\degr, $\Delta =
0\fdg46667$ is the beam separation, and $i_{b2}$ is 1 if $i_b$ is odd and 0 if
$i_b$ is even. The pointings within a cluster are defined by $(c_l,c_b) =$
(0,0), (1,0), (0,1) and ($-1$,1), and $n$ and $m$ are integers, the range of
which is determined by the area to be covered. For example, the
pointing closest to the Galactic Centre is at $l=359\fdg767$, $b=0\fdg0$,
with $i_l = 4999$ and $i_b = 500$, corresponding to $n=92$, $m=-23$ and
$(c_l,c_b) = (1,0)$. 

A record of the observational and processing status is maintained in a file,
where each pointing is identified by a 7-digit number, $1000\,i_l + i_b$,
known as the pointing ID. The inverse transformation, from $(l,b)$ to the nearest
pointing ID is given by 
\begin{eqnarray}
i_b & = & 500 + b/d_b +0.5~{\rm and} \\
i_l & = & 5000 + l/d_l - 0.5 i_{b2} + 0.5,
\end{eqnarray}
where $-180\degr < l \le 180\degr$. Each of the 13 beam positions has a
unique `grid ID' which, for a feed Galactic position angle of 30\degr, is
offset from the pointing ID by $\Delta i_l$ = 0, $-$1, 1, 2, 1, $-$1, $-$2, $-$3, 0,
3, 3, 0 and $-$3, and $\Delta i_b$ = 0, 2, 2, 0, $-$2, $-$2, 0, 2, 4, 2, $-$2, $-$4
and $-$2 respectively.

An interactive program, {\sc hexview}, is used to display the status of each
pointing and to select pointings for observation. Consecutive pointings
observed in one session are separated by about 5\degr~ to avoid the
possibility of a strong pulsar appearing in more than one pointing and hence
possibly being flagged as interference. As a system check, the strong pulsar
PSR J1359$-$6038 is observed on most observing days for about 1 min,
centred on each beam in turn.

Initially the survey region extended from $l=220\degr$. However, a
decision was made to limit it at $l=260\degr$ after a few months because of
the low pulsar density between these two longitudes.  Observations began at
low latitudes where the pulsar concentration is high. The discovery rate for
the first year of observation was at the unprecedented rate of more than one
pulsar per hour of observing time.

\section{TIMING OBSERVATIONS AND ANALYSIS}\label{sec:timing}
Almost all follow-up investigations require a more precise pulsar position,
pulsar period $P$, and/or period derivative $\dot P$ than those obtained
from the discovery observation. Improved estimates of the DM, the mean
pulsed flux density $S_{1400}$ and the pulse widths at the 50 per cent and
10 per cent levels, $W_{50}$ and $W_{10}$, are also valuable. All of these
parameters are determined from a series of timing observations made over a
span of at least one year. These observations also reveal binary motion if
present, and enable the binary parameters to be determined.

Timing observations are made using either the Parkes 64-m telescope or the
Lovell 76-m telescope at Jodrell Bank Observatory, with most of the detected
pulsars north of declination $-35\degr$ being timed at Jodrell Bank. In this
paper, we give results only from Parkes timing observations. The centre beam
of the multibeam receiver is used, with the same filterbank and data
acquisition system as is used for the survey. Typically, observations are of
duration between 2 and 30\,min, dependent upon the pulsar flux density, and
are made at intervals of 2 -- 6 weeks, with some more closely spaced
observations to resolve pulse counting ambiguities.

The data for each observation are dedispersed and synchronously folded at
the predicted topocentric pulsar period in off-line processing to form an
`archive' file. These files normally have 8 sub-bands across the observed
bandwidth and a series of sub-integrations, typically of 1-min
duration. These are summed over both frequency and time to form a mean pulse
profile. This is then convolved with a `standard profile' for the
corresponding pulsar, producing a topocentric time-of-arrival (TOA).  These
are then processed using the {\sc tempo} program\footnote{See
http://pulsar.princeton.edu/tempo or
http://www.atnf.csiro.au/research/pulsar/timing/tempo.} which converts them
to barycentric TOAs at infinite frequency and performs a multi-parameter fit
for the pulsar parameters.  Barycentric corrections are obtained using the
Jet Propulsion Laboratory DE200 solar-system ephemeris
\cite{sta90}. Initially, standard profiles are formed from a high
signal-to-noise ratio observation. Once a valid timing solution is obtained,
all or most of the observations are summed to form a `grand average'
profile. A new standard profile is then made from this average profile and
the TOAs recomputed. This often reduces the final residuals for the timing
solution by a factor of two or more.

As evidenced by the discovery that PSR J2144$-$3933 has an 8.5-s period
\cite{ymj99}, standard search software can sometimes mis-identify the pulse
period by a factor of two or three. As mentioned above (\S\ref{sec:sens})
there is a software limit at a period of 5 s. Furthermore,
interference can sometimes mask low-frequency spectral components. In such
cases a pulsar may be detected by its 2nd or 3rd harmonic, leading to the
assumption of an incorrect period. Such errors can be identified by folding
the data at twice and three times the nominal period and examining the
resulting mean pulse profiles. This check is routinely done for all pulsars
discovered in this survey and has resulted in period correction for several
pulsars.

In a few pulsars, at the confirmation stage or soon after, significant
variations in solar-system barycentric period are observed. These may be due
to an especially large period derivative, or to binary motion. In either
case, an improved estimate of the barycentric period is obtained by summing
the archive sub-integrations over a range of periods about the nominal
value. Where the rate of period change is not too great, improved periods
can be obtained by fitting TOAs for several observations over one or a few
adjacent days. A series of these barycentric periods can then be fitted with
either a period derivative term or a binary model. The parameters from this
fit then form the basis for a coherent timing solution using {\sc tempo}. 

Improved estimates of the dispersion measure can also be obtained from
individual archive files by summing the sub-bands with a range of delays
corresponding to different DM values about the nominal value and searching
for the highest signal-to-noise ratio. After a timing
solution is available, a final DM value for each pulsar is obtained by
summing each archive in time and forming four sub-bands across the 288 MHz
observed bandwidth. TOAs are then obtained for all archives for each of the
four sub-bands. Improved estimates of the DM and its error are then obtained
using {\sc tempo}, holding all parameters except DM fixed at the values from
the final timing solution.

The grand average profile for each pulsar is also used as a basis for
estimating the mean flux density and pulse width parameters. Flux densities
were calibrated by observing a sample of 13 pulsars with previously
catalogued 1400 MHz flux densities of moderate value (to give reasonable
signal-to-noise ratio while avoiding digitiser saturation) and high DMs (to
minimise variations due to scintillation).  Table~\ref{tbl:flux_cal} lists
the pulsars used, their DM and their assumed flux density \cite{tml93}. This
calibration is based on the accumulated digitiser counts with the multibeam
system, and hence is relative to the system equivalent flux density. The
effect of the varying sky background temperature was allowed for in the
calculation by scaling values of sky background temperature at 408 MHz from
the Haslam et al. (1982)\nocite{hssw82} all-sky survey to 1374 MHz assuming
a spectral index of $-2.5$. Based on the rms fluctuation of computed flux
densities among the calibration pulsars and independently calibrated
observations of these and other pulsars using the Australia Telescope
Compact Array and the Caltech correlator \cite{nms+97}, we estimate that the
flux scale is accurate at the 10 -- 15 per cent level.

\begin{table}
\caption{Flux density calibration pulsars}
\begin{tabular}{lrl}
\hline
PSR J & DM~~~ & $S_{1400}$ \\
  &  cm$^{-3}$ pc & mJy \\
\hline
1157$-$6224 &    325.2!! &  10  \\
1224$-$6407 &     97.8!! &  !5  \\
1243$-$6423 &    297.2!! &  13  \\
1306$-$6617 &    436.9!! &  !3.9  \\
1326$-$5859 &    288.1!! &  10  \\
1327$-$6222 &    318.4!! &  12  \\
1327$-$6301 &    294.9!! &  !3.4  \\
1338$-$6204 &    638.0!! &  !5.1  \\
1359$-$6038 &    294.1!! &  !7  \\
1430$-$6623 &     65.3!! &  !6  \\
1512$-$5759 &    628.7!! &  !4.0  \\
1522$-$5829 &    199.9!! &  !4.8  \\
1539$-$5626 &    176.5!! &  !4.2  \\
\hline
\end{tabular}
\label{tbl:flux_cal}
\end{table}

Except for a few especially interesting cases, timing observations cease 12
-- 18 months after confirmation. By this time a coherent timing solution has
normally been obtained, giving an accurate pulsar position, pulse period,
period derivative, dispersion measure and, if applicable, binary
parameters. Pulsars are renamed at this stage, based on the accurate J2000
position. The parameters are then entered into the pulsar catalogue, allowing
accurate predictions for future observations, and listed on the Parkes
multibeam pulsar survey {\tt New Pulsars} web page. The multibeam pulsar survey
web pages also specify policy for release of raw data tapes. On request,
these are made available for copying two years after the date of
recording. The {\tt Data Release} web page lists all available observations
sorted by date, Parkes project identification, observed position and tape
label. We will provide documentation specifying the data format and
software to read and copy data tapes on request.

\section{DISCOVERY AND TIMING OF THE FIRST 100 PULSARS}\label{sec:100psrs}
In this paper we report the discovery of 100 pulsars by the Parkes multibeam
pulsar survey. These pulsars were selected as the first 100 from the list of
pulsars being timed at Parkes, ordered by the date at which regular Parkes
timing observations commenced. All are south of declination
$-35\degr$. Table~\ref{tb:posn} lists the pulsar name, the J2000 right
ascension and declination from the timing solution, the corresponding
Galactic coordinates, the beam in which the pulsar was detected, the radial
distance of the pulsar from the beam centre in units of the beam radius
(cf. Table~\ref{tb:rcvr}), the signal-to-noise ratio of the discovery
observation from the final time-domain folding in the search process, the
mean flux density averaged over all observations included in the timing
solution, and pulse widths at 50 per cent and 10 per cent of the peak of the
mean pulse profile. Flux densities have been corrected for off-centre
pointing during the timing observations. Many of these pulsars were detected
more than once by the survey.  Beam and signal-to-noise details refer to the
detection having the highest signal-to-noise ratio. The 10 per cent width is
not measurable for pulsars with mean profiles having poor signal-to-noise
ratio. Estimated uncertainties are given in parentheses where relevant and
refer to the last quoted digit. Flux densities may be somewhat over
estimated for very weak pulsars or those which have extended null periods,
since non-detections are not included in the timing solution.

\begin{table*}
\begin{minipage}{150mm}
\caption{Positions, flux densities and widths for 100 pulsars discovered in
Parkes multibeam pulsar survey}
\begin{tabular}{lllrrccrllr}
\hline
\multicolumn{1}{c}{PSR J} & R.A. (J2000) & Dec. (J2000) & 
\multicolumn{1}{c}{$l$} & \multicolumn{1}{c}{$b$} & Beam & Radial &  
 \multicolumn{1}{c}{$S/N$} & \multicolumn{1}{c}{$S_{1400}$} & \multicolumn{1}{c}{$W_{50}$} & 
\multicolumn{1}{c}{$W_{10}$} \\
  & (h~~~m~~~s) & (~\degr ~~~\arcmin ~~~\arcsec) & 
\multicolumn{1}{c}{(\degr)} & \multicolumn{1}{c}{(\degr)} &   & Dist. &  &
\multicolumn{1}{c}{(mJy)} & \multicolumn{1}{c}{(ms)} &
\multicolumn{1}{c}{(ms)} \\ 
\hline
0835$-$3707  & 08:35:03.08(3)  & $-$37:07:51.5(3)  & 257.08 & +1.99 &   !2 & 1.86 & 20.7 &   0.28(4)&  !!4.9 &   12 \\
0838$-$3947  & 08:38:30.8(5)   & $-$39:47:22(7)    & 259.61 & +0.93 &   !6 & 0.86 & 11.5 &   0.11(2)&    !57 &    -- \\
0901$-$4624  & 09:01:40.12(3)  & $-$46:24:48.5(5)  & 267.40 & $-$0.00 & 11 & 1.52 &  9.6 &   0.46(6)&  !!5.1 &   36 \\
0922$-$4949  & 09:22:14.96(1)  & $-$49:49:12.08(8) & 272.24 & +0.16 &   !9 & 1.04 & 58.5 &   0.52(6)&  !!6.9 &   31 \\
0940$-$5428  & 09:40:58.22(4)  & $-$54:28:40.6(3)  & 277.51 & $-$1.29 & 12 & 0.93 & 33.3 &   0.35(4)&  !!9.6 &   -- \\ \\
0954$-$5430  & 09:54:06.04(3)  & $-$54:30:53.5(7)  & 279.00 & $-$0.10 & !6 & 0.90 & 60.9 &   0.36(5)&  !!7.5 &   20 \\
0957$-$5432  & 09:57:56.01(3)  & $-$54:32:03.9(5)  & 279.45 & +0.23 &   !5 & 0.62 & 20.0 &   0.18(3)&  !!3.6 &    8 \\
1001$-$5559  & 10:01:08.60(3)  & $-$55:59:00.2(3)  & 280.69 & $-$0.65 & !8 & 1.63 & 14.2 &   0.64(7)&  !17.0 &   59 \\
1002$-$5559  & 10:02:57.9(3)   & $-$55:59:37(5)    & 280.90 & $-$0.50 & !7 & 0.82 & 11.4 &   0.12(2)&    !25 &    -- \\
1016$-$5819  & 10:16:12.10(2)  & $-$58:19:01.15(8) & 283.71 & $-$1.36 & !7 & 1.14 & 15.9 &   0.31(4)&  !!4.2 &  -- \\ \\
1049$-$5833  & 10:49:50.34(9)  & $-$58:33:45(1)    & 287.63 & +0.65 &   12 & 0.45 & 53.0 &   0.72(8)&    !33 &   59 \\
1056$-$5709  & 10:56:43.8(1)   & $-$57:09:34(1)    & 287.84 & +2.31 &   10 & 0.62 & 11.7 &   0.11(2)&  !16.0 &    -- \\
1112$-$6103  & 11:12:14.81(4)  & $-$61:03:31.1(6)  & 291.22 & $-$0.46 & 13 & 0.47 & 42.7 &  1.40(15)&  !11.0 &    -- \\
1115$-$6052  & 11:15:53.68(3)  & $-$60:52:17.8(5)  & 291.56 & $-$0.13 & 11 & 0.46 & 28.7 &   0.38(5)&  !!5.4 &   10 \\
1119$-$6127  & 11:19:14.30(2)  & $-$61:27:49.5(2)  & 292.15 & $-$0.54 & !3 & 1.41 & 33.8 &   0.90(9)&    !24 &  48 \\ \\
1123$-$6102  & 11:23:41.70(6)  & $-$61:02:06.2(3)  & 292.51 & +0.05 &   !3 & 1.46 & 16.7 &   0.53(6)&  !10.0 &   25 \\
1130$-$5925  & 11:30:10.4(1)   & $-$59:25:34.1(7)  & 292.75 & +1.83 &   !4 & 0.43 & 17.2 &   0.12(2)&  !16.0 &    -- \\
1138$-$6207  & 11:38:21.62(3)  & $-$62:07:59.3(3)  & 294.51 & $-$0.46 & !1 & 0.82 & 23.0 &   0.49(6)&  !12.0 &    -- \\
1142$-$6230  & 11:42:52.5(3)   & $-$62:30:04(1)    & 295.11 & $-$0.68 & 12 & 0.68 & 15.8 &   0.26(4)&    !30 &    -- \\
1144$-$6146  & 11:44:34.8(3)   & $-$61:46:49(3)    & 295.12 & +0.07 &   11 & 0.57 & 63.3 &   0.45(6)&    !33 &   -- \\ \\
1144$-$6217  & 11:44:02.11(5)  & $-$62:17:30.3(4)  & 295.19 & $-$0.44 & !4 & 0.57 & 30.4 &   0.20(3)&  !10.0 &   27 \\
1216$-$6223  & 12:16:41.9(1)   & $-$62:23:57.8(9)  & 298.92 & +0.20 &   !1 & 0.54 & 13.2 &   0.15(3)&  !15.0 &    -- \\
1220$-$6318  & 12:20:17.9(1)   & $-$63:18:46(1)    & 299.44 & $-$0.65 & 10 & 0.68 & 32.0 &   0.68(8)&    !58 &    -- \\
1224$-$6208  & 12:24:44.25(8)  & $-$62:08:41.1(7)  & 299.82 & +0.57 &   12 & 0.97 & 16.2 &   0.23(3)&  !10.0 &   21 \\
1232$-$6501  & 12:32:17.840(5) & $-$65:01:03.33(4) & 300.91 & $-$2.22 & !7 & 0.51 & 23.6 &   0.34(4)&  !11.0 & 14 \\ \\
1245$-$6238  & 12:45:21.1(1)   & $-$62:38:55.9(8)  & 302.23 & +0.21 &   !9 & 0.18 & 16.2 &   0.14(2)&    !62 &    -- \\
1252$-$6314  & 12:52:42.6(1)   & $-$63:14:32.7(6)  & 303.08 & $-$0.37 & !8 & 1.89 & 24.9 &   0.66(8)&    !20 &   41 \\
1301$-$6305  & 13:01:45.8(1)   & $-$63:05:34(1)    & 304.10 & $-$0.24 & !8 & 0.21 & 18.6 &   0.46(6)&    !28 &    -- \\
1303$-$6305  & 13:03:00.0(2)   & $-$63:05:01(1)    & 304.24 & $-$0.24 & 13 & 0.96 & 26.6 &   0.36(5)&    !38 &   78 \\
1305$-$6203  & 13:05:20.9(3)   & $-$62:03:22(1)    & 304.56 & +0.77 &   12 & 0.78 & 31.2 &   0.62(7)&  !16.0 &   -- \\ \\
1305$-$6256  & 13:05:28.0(4)   & $-$62:56:39(3)    & 304.53 & $-$0.12 & !2 & 1.24 & 17.0 &   0.32(4)&  !19.0 &    -- \\
1307$-$6318  & 13:07:54.7(6)   & $-$63:18:35(4)    & 304.78 & $-$0.50 & 10 & 1.54 & 29.4 &  1.40(15)&   505 &    -- \\
1309$-$6415  & 13:09:16.6(7)   & $-$64:15:59(5)    & 304.87 & $-$1.46 & !2 & 0.72 & 16.3 &   0.21(3)&    !26 &    -- \\
1312$-$6400  & 13:12:07.2(1)   & $-$64:00:55.6(9)  & 305.20 & $-$1.23 & !3 & 0.45 & 59.4 &   0.75(8)&    !34 &   61 \\
1317$-$6302  & 13:17:44.69(7)  & $-$63:02:52.2(6)  & 305.91 & $-$0.33 & 11 & 0.78 & 47.1 &  0.99(11)&  !12.0 & -- \\ \\
1322$-$6241  & 13:22:32.1(1)   & $-$62:41:53.5(8)  & 306.49 & $-$0.04 & !3 & 1.49 & 29.6 &   0.37(5)&  !!8.7 &   19 \\
1327$-$6400  & 13:27:10.3(1)   & $-$64:00:13.1(6)  & 306.84 & $-$1.40 & !3 & 0.57 & 29.1 &   0.36(5)&  !13.0 &  120 \\
1341$-$6023  & 13:41:07.37(3)  & $-$60:23:34.7(5)  & 309.04 & +1.89 &   !1 & 1.64 & 90.4 &   0.63(7)&  !!9.2 &   19 \\
1345$-$6115  & 13:45:44.4(2)   & $-$61:15:31(2)    & 309.41 & +0.93 &   !3 & 1.00 & 51.0 &   0.59(7)&    !27 &   40 \\
1347$-$5947  & 13:47:19.38(4)  & $-$59:47:39.8(5)  & 309.91 & +2.32 &   11 & 1.09 & 31.4 &   0.67(8)&  !11.0 &   19 \\ \\
1348$-$6307  & 13:48:42.4(4)   & $-$63:07:04(4)    & 309.35 & $-$0.96 & !2 & 1.10 & 17.5 &   0.51(6)&    !79 &    -- \\
1349$-$6130  & 13:49:36.65(4)  & $-$61:30:17.1(4)  & 309.81 & +0.59 &   !5 & 1.62 & 13.6 &   0.58(7)&  !!6.2 &   14 \\
1406$-$6121  & 14:06:50.04(6)  & $-$61:21:27.9(6)  & 311.84 & +0.20 &   !9 & 0.50 & 17.1 &   0.36(5)&  !16.0 &    -- \\
1407$-$6048  & 14:07:58.6(1)   & $-$60:48:59(1)    & 312.13 & +0.68 &   !7 & 0.98 & 13.2 &   0.20(3)&    !21 &    -- \\
1407$-$6153  & 14:07:56.5(5)   & $-$61:53:59(6)    & 311.81 & $-$0.35 & !1 & 0.50 & 16.2 &   0.36(5)&    !57 & -- \\ \\
1412$-$6111  & 14:12:59.6(1)   & $-$61:11:30.5(7)  & 312.60 & +0.14 &   !3 & 1.03 & 27.8 &   0.44(5)&  !12.0 &   22 \\
1412$-$6145  & 14:12:07.69(5)  & $-$61:45:28.8(6)  & 312.32 & $-$0.37 & !2 & 0.50 & 30.4 &   0.47(6)&  !12.0 &    -- \\
1413$-$6222  & 14:13:05.47(8)  & $-$62:22:28(1)    & 312.24 & $-$0.99 & 10 & 0.62 & 50.7 &  0.96(11)&    !23 &    -- \\
1416$-$6037  & 14:16:30.6(2)   & $-$60:37:59.5(9)  & 313.18 & +0.53 &   10 & 0.34 & 63.1 &   0.70(8)&  !13.0 &   20 \\
1425$-$6210  & 14:25:07.7(3)   & $-$62:10:05(1)    & 313.63 & $-$1.26 & 13 & 1.16 & 10.6 &   0.19(3)&  !11.0 &    -- \\
\hline
\end{tabular}   
\label{tb:posn}
\end{minipage}
\end{table*} 

\addtocounter{table}{-1}
\begin{table*}
\begin{minipage}{150mm}
\caption{-- {\it continued}}
\begin{tabular}{lllrrccrllr}
\hline
\multicolumn{1}{c}{PSR J} & R.A. (J2000) & Dec. (J2000) & 
\multicolumn{1}{c}{$l$} & \multicolumn{1}{c}{$b$} & Beam & Radial &
\multicolumn{1}{c}{$S/N$} & \multicolumn{1}{c}{$S_{1400}$} & \multicolumn{1}{c}{$W_{50}$} & 
\multicolumn{1}{c}{$W_{10}$} \\
  & (h~~~m~~~s) & (~\degr ~~~\arcmin ~~~\arcsec) & 
\multicolumn{1}{c}{(\degr)} & \multicolumn{1}{c}{(\degr)} &   & Dist. &  &
\multicolumn{1}{c}{(mJy)} & \multicolumn{1}{c}{(ms)} & \multicolumn{1}{c}{(ms)} \\
\hline
1429$-$5935  & 14:29:25.9(1)   & $-$59:35:59(1)    & 315.05 & +0.95 &   !2 & 0.58 & 12.8 &   0.11(2)  &  !14.0 &    -- \\
1434$-$6029  & 14:34:39.1(3)   & $-$60:29:49(3)    & 315.31 & $-$0.13 & 12 & 0.60 & 11.0 &   0.14(2)  &  !19.0 &    -- \\
1435$-$6100  & 14:35:20.2765(4) & $-$61:00:57.956(6)& 315.19 & $-$0.64 & !8 & 1.25 &12.1 &   0.25(4)  & !!1.10 &    -- \\
1444$-$5941  & 14:44:46.5(3)   & $-$59:41:19(3)    & 316.79 & +0.10 &   !2 & 0.88 & 13.1 &   0.42(5)  &    !47 &   79 \\
1452$-$5851  & 14:52:52.58(7)  & $-$58:51:13(2)    & 318.09 & +0.40 &   11 & 0.52 & 19.5 &   0.24(3)  &  !11.0 &    -- \\ \\
1454$-$5846  & 14:54:10.908(2) & $-$58:46:34.74(3) & 318.27 & +0.39 &   13 & 0.33 & 12.5 &   0.24(3) &  !!2.9 &    5 \\
1513$-$5739  & 15:13:58.99(9)  & $-$57:39:01(1)    & 321.10 & +0.10 &   !8 & 1.28 & 20.4 &   0.77(9)  &    !21 &   33 \\
1530$-$5327  & 15:30:26.87(6)  & $-$53:27:56.3(7)  & 325.33 & +2.35 &   !2 & 1.17 & 30.0 &   0.59(7)  &  !14.0 &    -- \\
1536$-$5433  & 15:36:04.8(2)   & $-$54:33:15(4)    & 325.37 & +0.98 &   11 & 1.83 & 40.5 &  1.30(14)  &    !36 &   66 \\
1537$-$5645  & 15:37:51.0(3)   & $-$56:45:04(7)    & 324.28 & $-$0.94 & 11 & 0.68 &  26.4&  1.00(11)  &   !67 &    -- \\ \\
1538$-$5438  & 15:38:49.0(2)   & $-$54:38:17(3)    & 325.64 & +0.68 &   !6 & 0.79 &  9.6 &   0.24(3)  &  !11.0 &    -- \\
1540$-$5736  & 15:40:59.0(1)   & $-$57:36:57(3)    & 324.11 & $-$1.89 & 12 & 0.98 & 13.9 &   0.24(3)  &  !14.0 &   27 \\
1543$-$5459  & 15:43:56.25(7)  & $-$54:59:14(1)    & 326.02 & $-$0.04 & 12 & 0.72 & 28.4 &   0.62(7)  &  !15.0 &   37 \\
1548$-$5607  & 15:48:44.03(3)  & $-$56:07:33.9(5)  & 325.86 & $-$1.36 & 11 & 0.31 & 60.2 &  1.00(11)  &  !!7.5 &   19 \\
1558$-$5419  & 15:58:41.5(2)   & $-$54:19:26(5)    & 328.10 & $-$0.87 & !7 & 0.93 & 18.9 &   0.40(5)  &    !22 &    -- \\ \\
1601$-$5244  & 16:01:27.3(3)   & $-$52:44:09(3)    & 329.45 & +0.07 &   !1 & 0.62 & 14.8 &   0.13(2)  &    !62 &    -- \\
1601$-$5335  & 16:01:54.91(6)  & $-$53:35:43(1)    & 328.94 & $-$0.63 & !9 & 0.77 & 16.9 &   0.22(3)  &  !!7.4 &    -- \\
1605$-$5215  & 16:05:19.0(4)   & $-$52:15:48(5)    & 330.20 & +0.03 &   !9 & 1.10 & 11.1 &   0.22(3)  &    !26 &    -- \\
1607$-$5140  & 16:07:49.3(3)   & $-$51:40:16(4)    & 330.88 & +0.21 &   !4 & 0.67 & 11.3 &   0.26(4)  &    !20 &    -- \\
1609$-$5158  & 16:09:26.7(5)   & $-$51:58:18(9)    & 330.87 & $-$0.18 & !1 & 0.71 & 12.1 &   0.27(4)  &   100 &    -- \\ \\
1610$-$5006  & 16:10:44.30(9)  & $-$50:06:42(2)    & 332.28 & +1.05 &   !4 & 1.55 & 15.5 &  1.60(17)  &    !42 &   98 \\
1611$-$4949  & 16:11:46.6(1)   & $-$49:49:57(1)    & 332.59 & +1.14 &   !3 & 1.23 & 18.9 &   0.58(7)  &  !18.0 &    -- \\
1613$-$5211  & 16:13:42.5(1)   & $-$52:11:21(2)    & 331.20 & $-$0.78 & !8 & 0.97 & 15.5 &   0.29(4)  &  !14.0 &    -- \\
1613$-$5234  & 16:13:57.5(2)   & $-$52:34:17(3)    & 330.96 & $-$1.09 & 13 & 0.67 & 13.9 &   0.28(4)  &    !31 &    -- \\
1616$-$5109  & 16:16:30.9(5)   & $-$51:09:17(9)    & 332.23 & $-$0.34 & !2 & 0.55 & 32.8 &  1.20(13)  &   !220 &    -- \\ \\
1616$-$5208  & 16:16:23.4(4)   & $-$52:08:48(4)    & 331.52 & $-$1.04 & !8 & 1.02 & 17.6 &   0.44(5)  &    !43 &    -- \\
1621$-$5039  & 16:21:04.7(2)   & $-$50:39:49(2)    & 333.08 & $-$0.49 & !3 & 1.46 & 11.9 &   0.36(5)  &    !20 &   42 \\
1622$-$4802  & 16:22:47.2(1)   & $-$48:02:13(1)    & 335.14 & +1.17 &   !4 & 0.87 & 38.5 &  0.92(10)  &  !17.0 &    -- \\
1622$-$4944  & 16:22:37.5(3)   & $-$49:44:30(3)    & 333.91 & $-$0.01 & !5 & 0.38 & 32.7 &   0.52(6)  &    !34 &    -- \\
1623$-$4949  & 16:23:54.8(2)   & $-$49:49:04(3)    & 334.00 & $-$0.21 & 10 & 1.53 & 13.0 &   0.36(5)  &  !11.0 &   31 \\ \\
1625$-$4904  & 16:25:18.1(1)   & $-$49:04:34(2)    & 334.69 & +0.14 &   !7 & 0.53 & 18.6 &   0.20(3)  &  !13.0 &    -- \\
1626$-$4807  & 16:26:42.5(3)   & $-$48:07:54(4)    & 335.53 & +0.64 &   !3 & 0.82 & 11.2 &   0.37(5)  &    !57 &    -- \\
1628$-$4804  & 16:28:26.8(1)   & $-$48:04:59(3)    & 335.77 & +0.46 &   !3 & 0.57 & 58.4 &  1.00(11)  &    !43 &  255 \\
1632$-$4621  & 16:32:49.81(2)  & $-$46:21:48.6(9)  & 337.53 & +1.10 &   !7 & 0.94 & 55.7 &  0.90(10)  &  !18.0 &   35 \\
1632$-$4818  & 16:32:40.0(2)   & $-$48:18:49(6)    & 336.08 & $-$0.21 & 11 & 0.31 & 20.4 &   0.39(5)  &    !43 &    -- \\ \\
1649$-$4349  & 16:49:20.42(8)  & $-$43:49:22(1)    & 341.36 & +0.60 &   !2 & 1.56 & 30.0 &   0.75(8)  &    !26 &    -- \\
1649$-$4729  & 16:49:18.3(1)   & $-$47:29:53(5)    & 338.54 & $-$1.76 & !3 & 0.93 & 11.8 &   0.29(4)  &  !18.0 &    -- \\
1650$-$4502  & 16:50:32.30(6)  & $-$45:02:37(2)    & 340.56 & $-$0.35 & 12 & 0.69 & 26.6 &   0.35(4)  &  !!7.3 &   19 \\
1653$-$4249  & 16:53:40.22(5)  & $-$42:49:03(2)    & 342.64 & +0.63 &   !6 & 1.63 & 26.4 &  1.30(14)  &  !14.0 &   29 \\
1709$-$3841  & 17:09:16.0(2)   & $-$38:41:17(10)   & 347.71 & +0.83 &   !9 & 0.65 & 24.0 &   0.31(4)  &  !19.0 &    -- \\ \\
1715$-$3700  & 17:15:09.7(2)   & $-$37:00:04(14)   & 349.76 & +0.89 &   !6 & 0.73 & 14.8 &   0.37(5)  &   110 &    -- \\
1716$-$3720  & 17:16:11.36(6)  & $-$37:20:44(3)    & 349.60 & +0.52 &   12 & 0.73 & 29.0 &   0.41(5)  &  !14.0 &  105 \\
1718$-$3825  & 17:18:13.565(4) & $-$38:25:18.1(2)  & 348.95 & $-$0.43 & !7 & 1.45 & 14.4 &  1.30(14)  &  !!3.9 &   14 \\
1720$-$3659  & 17:20:01.976(9) & $-$36:59:06.5(4)  & 350.33 & +0.10 &   12 & 1.43 & 14.1 &   0.74(8)  &  !!7.5 &   17 \\
1723$-$3659  & 17:23:07.580(6) & $-$36:59:13.9(3)  & 350.68 & $-$0.41 & !7 & 0.12 &113.7 &  1.50(16)  &  !!7.8 &   35 \\ \\
1724$-$3505  & 17:24:47.9(2)   & $-$35:05:36(7)    & 352.44 & +0.38 &   !3 & 1.21 & 10.0 &   0.24(3)  &    !24 &    -- \\
1725$-$3546  & 17:25:42.2(3)   & $-$35:46:16(7)    & 351.98 & $-$0.15 & 11 & 0.50 & 30.0 &   0.61(7)  &    !33 &    -- \\
1726$-$3530  & 17:26:07.6(4)   & $-$35:30:05(15)   & 352.25 & $-$0.07 & !3 & 0.73 & 18.0 &   0.30(4)  &    !55 &    -- \\
1726$-$3635  & 17:26:49.61(3)  & $-$36:35:46(1)    & 351.42 & $-$0.80 & !3 & 0.46 & 21.8 &   0.29(4)  &  !!7.8 &   66 \\
1728$-$3733  & 17:28:46.2(2)   & $-$37:33:08(9)    & 350.84 & $-$1.66 & 11 & 0.61 & 19.9 &   0.19(3)  &  !!8.1 &   18 \\
\hline
\end{tabular}   
\end{minipage}
\end{table*}

Table~\ref{tb:prd} gives solar-system barycentric pulse periods, period
derivatives, epoch of the period, the number of TOAs in the timing solution,
the MJD range covered by the timing observations, the final rms timing
residual and the dispersion measure.

\begin{table*}
\begin{minipage}{150mm}
\caption{Period parameters and dispersion measures for 100 pulsars discovered in
Parkes multibeam pulsar survey}
\begin{tabular}{lllcccrl}
\hline
\multicolumn{1}{c}{PSR J} & \multicolumn{1}{c}{Period, $P$} & \multicolumn{1}{c}{$\dot P$} &  
Epoch & $N_{toa}$ & Data Span & \multicolumn{1}{c}{Residual} & \multicolumn{1}{c}{DM}  \\
  & \multicolumn{1}{c}{(s)} & \multicolumn{1}{c}{($10^{-15}$)} & 
(MJD) & & (MJD)  & \multicolumn{1}{c}{($\mu$s)}  & \multicolumn{1}{c}{(cm$^{-3}$ pc)} \\ 
\hline
0835$-$3707  & 0.541404373627(15)   & !!!9.778(9)    & 51137.0 &   27 & 50940--51333 & 218!! & !112.3(3)   \\
0838$-$3947  & 1.7039457055(9)      & !!!0.8(4)      & 51162.0 &   19 & 50941--51382 &6149!! & !219(11)    \\
0901$-$4624  & 0.441995130786(14)   & !!87.494(8)    & 51031.0 &   20 & 50849--51212 &  90!! & !198.8(3)   \\
0922$-$4949  & 0.950288537028(8)    & !!97.569(4)    & 51279.0 &   36 & 51086--51471 & 203!! & !237.1(3)   \\
0940$-$5428  & 0.087545204308(4)    & !!32.8683(10)  & 51091.0 &   38 & 50849--51333 &1027!! & !134.5(9)   \\ \\
0954$-$5430  & 0.472834279266(16)   & !!43.912(12)   & 51034.0 &   26 & 50849--51219 & 185!! & !200.3(4)   \\
0957$-$5432  & 0.203556697536(6)    & !!!1.947(4)    & 51035.0 &   24 & 50849--51219 & 126!! & !226.1(3)   \\
1001$-$5559  & 1.66117674023(4)     & !!!0.860(4)    & 51172.0 &   36 & 50852--51490 & 511!! & !159.3(9)   \\
1002$-$5559  & 0.7775009067(3)      & !!!1.57(19)    & 51035.0 &   17 & 50849--51219 & 490!! & !426(4)     \\ 
1016$-$5819  & 0.0878341561432(9)   & !!!0.6980(4)   & 51155.0 &   30 & 50940--51370 & 124!! & !252.1(4)   \\ \\
1049$-$5833  & 2.2023250770(3)      & !!!4.41(15)    & 51031.0 &   16 & 50849--51212 &1080!! & !446.8(15)  \\
1056$-$5709  & 0.67608189374(8)     & !!!0.576(14)   & 51216.0 &   26 & 50940--51490 & 513!! & !436.5(18)  \\
1112$-$6103  & 0.064961851894(3)    & !!31.4596(13)  & 51055.0 &   45 & 50849--51261 & 785!! & !599.1(7)   \\
1115$-$6052  & 0.259776659501(9)    & !!!7.235(5)    & 51031.0 &   24 & 50849--51212 & 156!! & !228.2(4)   \\
1119$-$6127  & 0.40774589995(5)     & 4021.782(9)    & 51485.0 &   15 & 51391--51578 & 137!! & !707(2)  \\ \\
1123$-$6102  & 0.64023374765(3)     & !!!6.460(8)    & 51155.0 &   30 & 50940--51370 & 209!! & !439.4(4)   \\
1130$-$5925  & 0.68098383242(6)     & !!!0.952(7)    & 51172.0 &   34 & 50851--51491 & 586!! & !264.4(16)  \\
1138$-$6207  & 0.117563794023(3)    & !!12.4784(5)   & 51171.0 &   39 & 50849--51491 & 488!! & !519.8(8)   \\
1142$-$6230  & 0.55838338569(10)    & !!!0.08(4)     & 51137.0 &   16 & 50940--51334 & 347!! & !343.8(17)  \\
1144$-$6146  & 0.98778306934(16)    & !!$-$0.04(9)   & 51032.0 &   11 & 50851--51211 & 233!! & !78.7(13)   \\ \\
1144$-$6217  & 0.85066494337(4)     & !!30.835(7)    & 51110.0 &   28 & 50849--51370 & 439!! & !284.7(6)   \\
1216$-$6223  & 0.37404678583(5)     & !!16.819(7)    & 51111.0 &   22 & 50851--51369 & 224!! & !786.6(18)  \\
1220$-$6318  & 0.78921201205(8)     & !!!0.080(12)   & 51216.0 &   25 & 50940--51491 &1151!! & !347(3)     \\
1224$-$6208  & 0.58576120812(4)     & !!20.196(11)   & 51111.0 &   24 & 50851--51369 & 312!! & !454.2(7)   \\
1232$-$6501  & 0.0882819082341(3)   & !!!0.00081(2)  & 51270.0 &   72 & 50940--51856 & 200!! & !239.4(5)   \\ \\
1245$-$6238  & 2.2830933508(3)      & !!10.92(5)     & 51206.0 &   21 & 50941--51470 & 955!! & !336(2)     \\
1252$-$6314  & 0.82333927128(6)     & !!!0.11(3)     & 51155.0 &   25 & 50940--51369 & 675!! & !278.4(13)  \\
1301$-$6305  & 0.18452809509(6)     & !266.747(3)    & 51206.0 &   37 & 50940--51471 &1540!! & !374(3)     \\
1303$-$6305  & 2.3066415539(4)      & !!!2.18(16)    & 51138.0 &   16 & 50940--51335 & 475!! & !343(3)     \\
1305$-$6203  & 0.42776184224(8)     & !!32.14(3)     & 51138.0 &   17 & 50940--51335 & 230!! & !470.0(15)  \\ \\
1305$-$6256  & 0.47823093284(12)    & !!!2.11(4)     & 51138.0 &   15 & 50941--51335 & 202!! & !967(3)     \\
1307$-$6318  & 4.9624272525(20)     & !!21.1(4)      & 51206.0 &   20 & 50940--51471 &4348!! & !374(8)     \\
1309$-$6415  & 0.6194535568(3)      & !!!8.79(12)    & 51303.0 &   15 & 51087--51517 & 184!! & !574(5)     \\
1312$-$6400  & 2.43743249609(11)    & !!!0.68(5)     & 51303.0 &   26 & 51087--51522 & 974!! & !93.0(12)   \\
1317$-$6302  & 0.26127055606(3)     & !!!0.102(6)    & 51138.0 &   23 & 50940--51335 & 205!! & !678.1(12)  \\ \\
1322$-$6241  & 0.50605841373(5)     & !!!2.587(18)   & 51138.0 &   18 & 50940--51335 & 298!! & !618.8(19)  \\
1327$-$6400  & 0.280677974168(13)   & !!31.177(4)    & 51206.0 &   24 & 50940--51471 & 703!! & !680.9(14)  \\
1341$-$6023  & 0.627285365870(16)   & !!19.461(8)    & 51280.0 &   17 & 51088--51471 & 210!! & !364.6(9)   \\
1345$-$6115  & 1.25308459010(18)    & !!!3.25(8)     & 51138.0 &   17 & 50940--51335 & 366!! & !278(2)     \\
1347$-$5947  & 0.609961754304(15)   & !!14.160(7)    & 51294.0 &   14 & 51088--51500 & 252!! & !293.4(5)   \\ \\
1348$-$6307  & 0.9277722389(3)      & !!!3.79(7)     & 51304.0 &   19 & 51088--51522 &1039!! & !597(3)     \\
1349$-$6130  & 0.259362860073(9)    & !!!5.125(4)    & 51138.0 &   23 & 50940--51335 & 123!! & !284.6(4)   \\
1406$-$6121  & 0.213074653776(14)   & !!54.701(3)    & 51111.0 &   30 & 50851--51370 &1267!! & !542.3(18)  \\
1407$-$6048  & 0.49234420664(5)     & !!!3.156(8)    & 51161.0 &   23 & 50849--51471 & 926!! & !575.2(17)  \\
1407$-$6153  & 0.7016149492(3)      & !!!8.85(7)     & 51093.0 &   17 & 50851--51333 &2588!! & !645(9)     \\ \\
1412$-$6111  & 0.52915639797(4)     & !!!1.91(3)     & 51031.0 &   20 & 50849--51212 & 265!! & !311.8(9)   \\
1412$-$6145  & 0.315224970657(12)   & !!98.6598(13)  & 51186.0 &   40 & 50849--51522 & 612!! & !514.7(11)  \\
1413$-$6222  & 0.29240770249(3)     & !!!2.229(6)    & 51092.0 &   27 & 50849--51333 & 368!! & !808.1(12)  \\
1416$-$6037  & 0.29558048193(3)     & !!!4.280(15)   & 51031.0 &   18 & 50849--51212 & 169!! & !289.2(10)  \\
1425$-$6210  & 0.50173030987(8)     & !!!0.48(4)     & 51031.0 &   15 & 50849--51212 & 256!! & !430.1(17)  \\
\hline
\end{tabular}   
\label{tb:prd}
\end{minipage}
\end{table*}

\addtocounter{table}{-1}
\begin{table*}
\begin{minipage}{150mm}
\caption{-- {\it continued}}
\begin{tabular}{lllcccrl}
\hline
\multicolumn{1}{c}{PSR J} & \multicolumn{1}{c}{Period, $P$} & \multicolumn{1}{c}{$\dot P$} &  
Epoch & $N_{toa}$ & Data Span & \multicolumn{1}{c}{Residual} & \multicolumn{1}{c}{DM}  \\
  & \multicolumn{1}{c}{(s)} & \multicolumn{1}{c}{($10^{-15}$)} & 
(MJD) & & (MJD)  & \multicolumn{1}{c}{($\mu$s)}  & \multicolumn{1}{c}{(cm$^{-3}$ pc)} \\ 
\hline
1429$-$5935  & 0.76391483053(8)     & !!42.751(9)    & 51232.0 &  21 & 50940--51523 & 550!! & !457(2)     \\
1434$-$6029  & 0.96334832315(16)    & !!!1.03(8)     & 51137.0 &  16 & 50940--51333 & 407!! & !282(3)     \\
1435$-$6100  & 0.009347972210248(6) & !!!0.0000245(4)& 51270.0 &  93 & 50939--51856 &  14!! & !113.7(6)   \\
1444$-$5941  & 2.7602279448(6)      & !!!8.2(3)      & 51137.0 &  17 & 50941--51333 & 746!! & !177.1(19)  \\
1452$-$5851  & 0.38662501748(3)     & !!50.706(18)   & 51280.0 &  16 & 51088--51472 & 146!! & !262.4(15)  \\ \\
1454$-$5846  & 0.04524877299802(9)  & !!!0.000816(7) & 51300.0 &  81 & 50981--51856 & 100!! & !115.95(16) \\
1513$-$5739  & 0.97345803480(9)     & !!27.55(4)     & 51137.0 &  20 & 50941--51333 & 480!! & !469.7(10)  \\
1530$-$5327  & 0.278956721152(15)   & !!!4.683(4)    & 51253.0 &  26 & 51013--51491 & 402!! & !49.6(10)   \\
1536$-$5433  & 0.8814384311(3)      & !!!1.91(8)     & 51138.0 &  21 & 50941--51334 &1529!! & !147.5(19)  \\
1537$-$5645  & 0.43046412386(15)    & !!!2.78(6)     & 51306.0 &  18 & 51088--51524 & 830!! & !707(5)     \\ \\
1538$-$5438  & 0.27672613726(5)     & !!!1.42(3)     & 51138.0 &  19 & 50941--51334 & 885!! & !136.9(17)  \\ 
1540$-$5736  & 0.61291628569(8)     & !!!0.42(3)     & 51309.0 &  21 & 51089--51528 & 482!! & !304.5(13)  \\
1543$-$5459  & 0.37711856263(3)     & !!52.018(9)    & 51139.0 &  21 & 50941--51371 & 253!! & !345.7(10)  \\
1548$-$5607  & 0.170933992695(5)    & !!10.736(3)    & 51138.0 &  33 & 50941--51334 & 220!! & !315.5(4)   \\
1558$-$5419  & 0.59457526355(15)    & !!!6.04(6)     & 51138.0 &  16 & 50941--51334 & 480!! & !657(3)     \\ \\
1601$-$5244  & 2.559356631(7)       & !!!0.72(14)    & 51071.0 &  17 & 50849--51292 & 726!! & !273(3)     \\ 
1601$-$5335  & 0.288456511543(16)   & !!62.371(6)    & 51156.0 &  26 & 50941--51371 & 462!! & !194.6(7)   \\
1605$-$5215  & 1.0136087473(5)      & !!!4.75(11)    & 51072.0 &  20 & 50851--51292 &6570!! & !532(4)     \\
1607$-$5140  & 0.34272279247(9)     & !!!2.54(4)     & 51072.0 &  20 & 50849--51293 &2204!! & !533(3)     \\
1609$-$5158  & 1.2794023539(7)      & !!12.96(10)    & 51160.0 &  23 & 50849--51470 &3299!! & 1069(8)     \\ \\
1610$-$5006  & 0.48111885215(5)     & !!13.625(10)   & 51111.0 &  27 & 50849--51371 & 917!! & !416(3)     \\ 
1611$-$4949  & 0.66643792285(8)     & !!!0.54(3)     & 51071.0 &  22 & 50849--51292 & 546!! & !556.8(18)  \\
1613$-$5211  & 0.45750181782(7)     & !!19.231(17)   & 51072.0 &  16 & 50849--51293 & 259!! & !360(2)     \\
1613$-$5234  & 0.65522059567(18)    & !!!6.629(18)   & 51111.0 &  22 & 50849--51463 & 501!! & !624(4)     \\
1616$-$5109  & 1.2195938825(8)      & !!19.08(9)     & 51161.0 &  24 & 50849--51471 &3230!! & 1160(15)    \\ \\
1616$-$5208  & 1.0258308926(4)      & !!28.91(10)    & 51072.0 &  16 & 50849--51293 & 688!! & !488(3)     \\ 
1621$-$5039  & 1.08401994353(16)    & !!13.03(5)     & 51072.0 &  20 & 50849--51293 & 642!! & !261(3)     \\
1622$-$4802  & 0.26507223569(3)     & !!!0.307(15)   & 51033.0 &  17 & 50851--51213 & 160!! & !364.3(13)  \\
1622$-$4944  & 1.0729678942(3)      & !!17.08(7)     & 51072.0 &  21 & 50849--51293 &1711!! & !755(4)     \\
1623$-$4949  & 0.72573215540(9)     & !!42.09(4)     & 51073.0 &  21 & 50851--51293 & 374!! & !183.3(10)  \\ \\
1625$-$4904  & 0.46033949229(7)     & !!16.76(2)     & 51073.0 &  17 & 50852--51293 & 275!! & !684.8(17)  \\ 
1626$-$4807  & 0.29392818864(7)     & !!17.476(12)   & 51207.0 &  23 & 50941--51472 &5286!! & !817(6)     \\
1628$-$4804  & 0.86597096270(12)    & !!!1.24(4)     & 51073.0 &  20 & 50851--51293 & 643!! & !952(3)     \\
1632$-$4621  & 1.70915449528(4)     & !!76.02(3)     & 51291.0 &  18 & 51089--51492 & 340!! & !562.9(8)   \\
1632$-$4818  & 0.81342978867(20)    & !650.64(4)     & 51112.0 &  23 & 50852--51371 &1439!! & !758(5)     \\ \\
1649$-$4349  & 0.87071155949(7)     & !!!0.044(19)   & 51243.0 &  19 & 51013--51472 & 402!! & !398.6(12)  \\ 
1649$-$4729  & 0.29769219971(5)     & !!!6.550(16)   & 51157.0 &  19 & 50941--51371 &1280!! & !540.2(18)  \\
1650$-$4502  & 0.38086979928(5)     & !!16.061(16)   & 51118.0 &  15 & 50941--51294 & 228!! & !319.7(8)   \\
1653$-$4249  & 0.61255824122(5)     & !!!4.81(3)     & 51118.0 &  18 & 50940--51294 & 192!! & !416.1(11)  \\
1709$-$3841  & 0.58698616319(18)    & !!!7.86(11)    & 51117.0 &  17 & 50941--51292 & 995!! & !356(3)     \\ \\
1715$-$3700  & 0.7796281140(4)      & !!!0.15(5)     & 51118.0 &  23 & 50852--51383 &2975!! & !449(11)    \\ 
1716$-$3720  & 0.63031371285(6)     & !!17.970(12)   & 51112.0 &  25 & 50852--51371 & 418!! & !682.7(17)  \\
1718$-$3825  & 0.0746699205656(4)   & !!13.22167(7)  & 51184.0 &  37 & 50877--51490 &  52!! & !247.4(3)   \\
1720$-$3659  & 0.351124633722(5)    & !!!0.0327(9)   & 51111.0 &  30 & 50851--51369 & 162!! & !381.6(5)   \\
1723$-$3659  & 0.2027219378604(17)  & !!!8.0075(4)   & 51118.0 &  25 & 50851--51383 & 129!! & !254.2(3)   \\ \\
1724$-$3505  & 1.2217076921(3)      & !!21.10(9)     & 51074.0 &  20 & 50852--51294 &1251!! & !875(3)     \\ 
1725$-$3546  & 1.0324711993(4)      & !!15.00(8)     & 51074.0 &  17 & 50852--51294 &2137!! & !744(4)     \\
1726$-$3530  & 1.1100937711(12)     & 1217.94(5)     & 51154.0 &  27 & 50852--51455 &2577!! & !727(7)     \\
1726$-$3635  & 0.287431567184(13)   & !!!1.440(3)    & 51111.0 &  22 & 50852--51369 & 153!! & !539.2(7)   \\
1728$-$3733  & 0.61553824309(16)    & !!!0.07(4)     & 51112.0 &  19 & 50852--51370 & 257!! & !281.5(7)   \\
\hline
\end{tabular}
\end{minipage}
\end{table*}

Three of the pulsars in Tables \ref{tb:posn} and \ref{tb:prd} are members of
binary systems. As mentioned in \S\ref{sec:intro}, all three of these
pulsars have been previously published by Camilo et
al. (2001)\nocite{clm+01}; details are repeated here for
completeness. Table~\ref{tb:binpar} gives the binary parameters for these
pulsars obtained from the timing solutions. Two of these pulsars are in
low-eccentricity orbits, for which the longitude and time of periastron are
not well determined. For these pulsars the reference epoch is the time of
passage through the ascending node. PSR J1454$-$5846 has a larger
(although still small) eccentricity and the longitude and epoch of periastron
could be determined with precision.

\begin{table*}
\begin{minipage}{150mm}
\caption{Binary pulsar parameters (Camilo et al. 2001)}
\begin{tabular}{llll}
\hline
 & PSR J1232$-$6501 & PSR J1435$-$6100 & PSR J1454$-$5846 \\
\hline
Orbital period (d)      & 1.86327241(8) & 1.354885217(2) & 12.4230655(2) \\
Projected semi-major axis (s) & 1.61402(6) & 6.184023(4)  & 26.52890(4) \\
Eccentricity     & 0.00011(8) & 0.000010(2) & 0.001898(3) \\
Longitude of periastron (deg) & 129(45) & 10(6) & 310.1(1) \\
Epoch of ascending node (MJD) & 51269.98417(2) & 51270.6084449(5) & 51303.833(4)$^*$ \\
\hline
$^*$ Epoch of periastron
\end{tabular}  
\label{tb:binpar}
\end{minipage}
\end{table*}

Mean pulse profiles at 1374 MHz for the 100 pulsars are given in
Fig.~\ref{fg:prf}. As mentioned in \S\ref{sec:timing}, these profiles were
formed by adding all data used for the timing solution. They typically have
several hours of effective integration time. For display purposes, these profiles
have been corrected for the effects of the high-pass filter in the
digitiser. To apply this correction, the profile is first given zero
mean. The corrected profile $b_n$, where $n$ is the bin number and $N$ is the
number of bins in the profile, is then given by
\begin{eqnarray}
b_n & = & a_n, \;\;\; (n=0) \nonumber\\
b_n & = & a_n + (t_{bin}/\tau_{\rm HP}) \sum_{m=0}^{n-1} a_{m}, \;\;\;(0<n<N)
\end{eqnarray}
where $a_n$ is the uncorrected zero-mean profile, $t_{bin}$ is the length of
each profile bin in seconds and $\tau_{\rm HP}$ is the high-pass filter time
constant in seconds. The value of $\tau_{\rm HP} = 0.9$~s was empirically
determined by requiring a flat corrected baseline on several long-period
pulsars.

\begin{figure} 
\centerline{\psfig{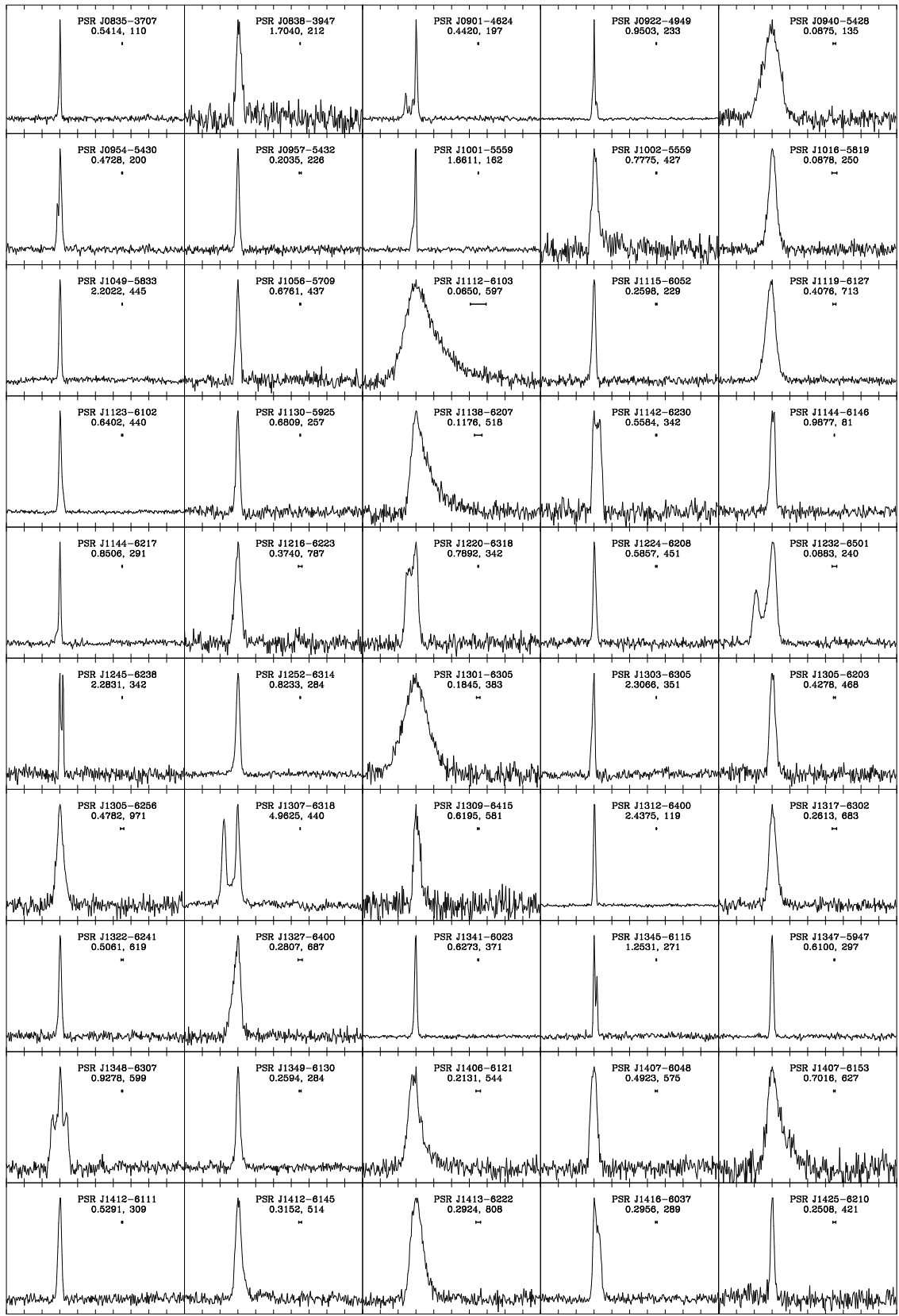}} 
\caption{Mean pulse profiles for 100 pulsars discovered in the Parkes
multibeam survey. The highest point in the profile is placed at phase
0.3. For each profile, the pulsar name, pulse period (in seconds) and DM
(in cm$^{-3}$ pc) are given. The small horizontal bar under the period
indicates the effective resolution of the profile, including the effects of
interstellar dispersion.}
\label{fg:prf}
\end{figure}
\addtocounter{figure}{-1}
\begin{figure} 
\centerline{\psfig{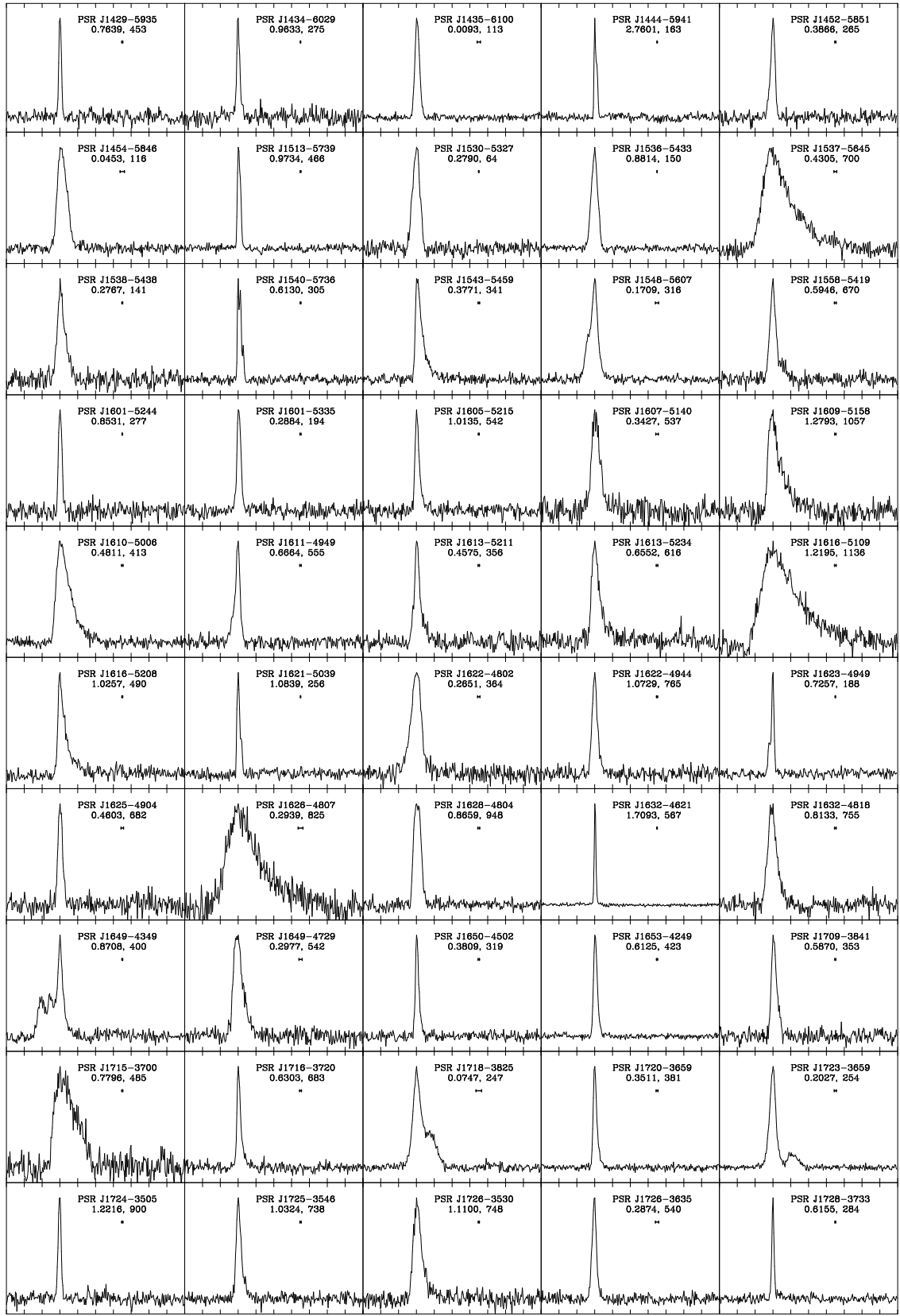}} 
\caption{-- {\it continued}}
\end{figure}

Prior to the commencement of the Parkes multibeam survey, there were 731
known radio pulsars, of which 693 are in the Galactic disk. (Five are in the
Magellanic Clouds and 33 are in globular clusters.) Of the 693 disk pulsars,
247 lie within the nominal search area of the multibeam survey. Since the
current survey is much more sensitive than any previous survey of this
region, we would expect to redetect essentially all of these pulsars. Because of the
current incompleteness of the survey, a definitive list of detected
previously known pulsars is deferred to a later paper.

\section{DISCUSSION AND CONCLUSIONS}
In this paper we have described in some detail the Parkes multibeam pulsar
survey, currently being conducted using a 13-beam receiver operating at a
central frequency of 1374 MHz on the Parkes 64-m radio telescope. Data
acquisition and analysis techniques are described and a detailed discussion
of the survey sensitivity and observing strategy is given. After
confirmation of a candidate, timing data are obtained, typically over a 12
-- 18 month period, giving an accurate position, pulse period, period
derivative and DM. The pulse width and mean flux density are estimated from
the mean pulse profile. We give the principal observed properties of the
first 100 pulsars discovered in the survey.

Table~\ref{tb:deriv} gives derived parameters for these 100 pulsars. After
the name, the first three columns give the log$_{10}$ of the characteristic
age, $\tau_c = P/(2\dot P)$, in years, the surface dipole magnetic field,
$B_s = 3.2 \times 10^{19} (P \dot P)^{1/2}$, in Gauss, and the rate of loss
of rotational energy, $\dot E = 4\pi^2 I \dot P P^{-3}$, in erg s$^{-1}$,
where a neutron-star moment of inertia $I = 10^{45}$ g cm$^2$ is
assumed. The next two columns give the pulsar distance, $d$, computed from
the DM assuming the Taylor \& Cordes (1993)\nocite{tc93} model for the
Galactic distribution of free electrons, and the implied Galactic
$z$-distance. Although distances are quoted to 0.1 kpc, in fact they are
generally more uncertain than that owing to uncertainties in the electron
density model. This is especially so for pulsars with very large DMs,
indicating large distances from the Sun. The final column gives the radio
luminosity $L_{1400} = S_{1400} d^2$. Pulsars discovered at relatively high
radio frequencies, for example, at 1400 MHz, tend to have a flatter spectrum
than those discovered at lower frequencies. For example, the sample of
pulsars discovered by Johnston et al. (1992)\nocite{jlm+92} has a mean
spectral index of $-1.0$ compared to the value of $-1.7$ found for pulsars
detected in the Parkes 70-cm survey\cite{tbms98}. However, the Johnston et
al. and Clifton et al. surveys were the first extensive surveys at these
higher frequencies. Most of the previously discovered pulsars had been found
in lower-frequency searches, which selected the steeper-spectrum
pulsars. The present survey is much more sensitive than any previous survey
of this region, and hence the discovered pulsars are a largely unbiased
sample. Adopting a compromise mean spectral index of $-1.3$ for the
multibeam discoveries, the $L_{1400}$ values may be converted to the more
commonly quoted 400 MHz luminosity by multiplying by 5.0. 

\begin{table*}
\begin{minipage}{150mm}
\caption{Derived parameters for 100 pulsars discovered in Parkes multibeam pulsar survey}
\label{tb:deriv}
\begin{tabular}{lcccccc}
\hline
\multicolumn{1}{c}{PSR J} & log[$\tau_c$ (yr)] & log[$B_s$ (G)] & log[$\dot E$ (erg s$^{-1}$)] 
& Distance & $z$ & $L_{1400}$ \\
 & & & & (kpc) & (kpc) & (mJy kpc$^2$) \\ 
\hline
0835$-$3707  &   5.94 &  12.37 &  33.38 & !2.3  &  +0.08 &  !!1.5    \\
0838$-$3947  &   7.52 &  12.08 &  30.81 & !8.2  &  +0.13 &  !!7.4    \\
0901$-$4624  &   4.90 &  12.80 &  34.60 & !7.5  &  $-$0.00 &  !25.6    \\
0922$-$4949  &   5.19 &  12.99 &  33.65 & 10.4  &  +0.03 &  !56.2    \\
0940$-$5428  &   4.63 &  12.24 &  36.28 & !4.3  &  $-$0.10 &  !!6.4    \\ \\
0954$-$5430  &   5.23 &  12.66 &  34.20 & !6.2  &  $-$0.01 &  !13.8    \\
0957$-$5432  &   6.22 &  11.80 &  33.96 & !7.0  &  +0.03 &  !!8.8    \\
1001$-$5559  &   7.49 &  12.08 &  30.87 & !3.9  &  $-$0.04 &  !!9.9    \\
1002$-$5559  &   6.89 &  12.05 &  32.11 & 17.4  &  $-$0.15 &  !36.3    \\
1016$-$5819  &   6.30 &  11.40 &  34.61 & !4.6  &  $-$0.11 &  !!6.6    \\ \\
1049$-$5833  &   6.90 &  12.50 &  31.20 & !9.7  &  +0.11 &  !68.0    \\
1056$-$5709  &   7.27 &  11.80 &  31.87 & 17.6  &  +0.71 &  !34.1    \\
1112$-$6103  &   4.51 &  12.16 &  36.65 &$>30.0$&  $<-$0.24 & $>1260.0$   \\
1115$-$6052  &   5.76 &  12.14 &  34.20 & !6.8  &  $-$0.01 &  !17.4    \\
1119$-$6127  &   3.21 &  13.61 &  36.36 &$>30.0$&  $<-$0.28 & $>720.0$    \\ \\
1123$-$6102  &   6.20 &  12.31 &  32.99 & 14.7  &  +0.01 &  114.5    \\
1130$-$5925  &   7.05 &  11.91 &  32.08 & !8.3  &  +0.27 &  !!8.2    \\
1138$-$6207  &   5.17 &  12.09 &  35.48 & 24.5  &  $-$0.20 &  294.1    \\
1142$-$6230  &   8.05 &  11.32 &  31.26 & 10.8  &  $-$0.13 &  !30.3    \\
1144$-$6146  & $>8.50$&$<11.35$&$<31.35$& 10.8  &  $-$0.13 &  !30.3    \\ \\
1144$-$6217  &   5.64 &  12.71 &  33.30 & !8.9  &  $-$0.07 &  !15.8    \\
1216$-$6223  &   5.55 &  12.40 &  34.11 &$>30.0$&  $>+0.10$ &  $>135.0$    \\
1220$-$6318  &   8.19 &  11.41 &  30.81 & 14.0  &  $-$0.16 &  133.3    \\
1224$-$6208  &   5.66 &  12.54 &  33.60 & 23.6  &  +0.23 &  128.1    \\
1232$-$6501  &   9.25 &   9.93 &  31.66 & 10.0  &  $-$0.39 &  !30.0    \\ \\
1245$-$6238  &   6.52 &  12.70 &  31.56 & 14.6  &  +0.05 &  !29.8    \\
1252$-$6314  &   8.08 &  11.48 &  30.89 & 11.0  &  $-$0.07 &  !79.9    \\
1301$-$6305  &   4.04 &  12.85 &  36.23 & 15.8  &  $-$0.07 &  114.8    \\
1303$-$6305  &   7.22 &  12.36 &  30.85 & 13.6  &  $-$0.06 &  !66.6    \\
1305$-$6203  &   5.32 &  12.57 &  34.20 & 24.1  &  +0.32 &  360.1    \\ \\
1305$-$6256  &   6.55 &  12.01 &  32.88 &$>30.0$&  $<-$0.06 &  $>288.0$    \\
1307$-$6318  &   6.57 &  13.02 &  30.83 & 14.4  &  $-$0.12 &  290.3    \\
1309$-$6415  &   6.05 &  12.37 &  33.18 &$>30.0$&  $<-$0.77 &  $>189.0$    \\
1312$-$6400  &   7.75 &  12.12 &  30.28 & !2.2  &  $-$0.05 &  !!3.7    \\
1317$-$6302  &   7.61 &  11.22 &  32.36 &$>30.0$&  $<-$0.17 &  $>891.0$    \\ \\
1322$-$6241  &   6.49 &  12.06 &  32.90 & 19.9  &  $-$0.01 &  146.5    \\
1327$-$6400  &   5.15 &  12.48 &  34.75 &$>30.0$&  $<-$0.73 &  $>324.0$    \\
1341$-$6023  &   5.71 &  12.55 &  33.49 & !7.0  &  +0.23 &  !31.1    \\
1345$-$6115  &   6.79 &  12.31 &  31.81 & !5.9  &  +0.10 &  !20.3    \\
1347$-$5947  &   5.83 &  12.47 &  33.40 & !6.5  &  +0.26 &  !28.6    \\ \\
1348$-$6307  &   6.59 &  12.28 &  32.28 & !8.2  &  $-$0.14 &  !34.1    \\
1349$-$6130  &   5.90 &  12.07 &  34.08 & !5.8  &  +0.06 &  !19.6    \\
1406$-$6121  &   4.79 &  12.54 &  35.34 & !9.1  &  +0.03 &  !29.9    \\
1407$-$6048  &   6.39 &  12.10 &  33.00 & !9.7  &  +0.12 &  !18.8    \\
1407$-$6153  &   6.10 &  12.40 &  33.00 & !9.8  &  $-$0.06 &  !34.2    \\ \\
1412$-$6111  &   6.64 &  12.01 &  32.71 & !6.0  &  +0.01 &  !15.6    \\
1412$-$6145  &   4.70 &  12.75 &  35.08 & !9.3  &  $-$0.06 &  !40.8    \\
1413$-$6222  &   6.32 &  11.91 &  33.54 & 27.7  &  $-$0.48 &  736.6    \\
1416$-$6037  &   6.04 &  12.06 &  33.81 & !5.7  &  +0.05 &  !22.9    \\
1425$-$6210  &   7.22 &  11.70 &  32.18 & 10.0  &  $-$0.22 &  !18.9    \\
\hline
\end{tabular}
\end{minipage}
\end{table*}

\addtocounter{table}{-1} 
\begin{table*}
\begin{minipage}{150mm}
\caption{-- {\it continued}}
\begin{tabular}{lcccccc}
\hline
\multicolumn{1}{c}{PSR J} & log[$\tau_c$ (yr)] & log[$B_s$ (G)] & log[$\dot E$ (erg s$^{-1}$)] 
& Distance & $z$ & $L_{1400}$ \\
 & & & & (kpc) & (kpc) & (mJy kpc$^2$) \\ 
\hline
1429$-$5935  &   5.45 &  12.76 &  33.58 & 10.6  &  +0.18 &  !12.4    \\
1434$-$6029  &   7.17 &  12.00 &  31.65 & !5.8  &  $-$0.01 &  !!4.7    \\
1435$-$6100  &   9.81 &   8.67 &  33.04 & !3.2  &  $-$0.04 &  !!2.1    \\
1444$-$5941  &   6.73 &  12.68 &  31.18 & !4.4  &  +0.01 &  !!8.1    \\
1452$-$5851  &   5.08 &  12.65 &  34.54 & !5.6  &  +0.04 &  !!7.6    \\ \\
1454$-$5846  &   8.95 &   9.78 &  32.53 & !3.3  &  +0.02 &  !!2.2    \\
1513$-$5739  &   5.75 &  12.72 &  33.08 & !9.8  &  +0.02 &  !74.6    \\
1530$-$5327  &   5.97 &  12.06 &  33.93 & !1.5  &  +0.06 &  !!1.3    \\
1536$-$5433  &   6.86 &  12.12 &  32.04 & !3.7  &  +0.06 &  !18.0    \\
1537$-$5645  &   6.39 &  12.05 &  33.15 & 24.7  &  $-$0.41 &  610.1    \\ \\
1538$-$5438  &   6.49 &  11.80 &  33.43 & !3.6  &  +0.04 &  !!3.1    \\
1540$-$5736  &   7.36 &  11.71 &  31.86 & !8.2  &  $-$0.27 &  !16.0    \\
1543$-$5459  &   5.06 &  12.65 &  34.58 & !6.3  &  $-$0.00 &  !24.8    \\
1548$-$5607  &   5.40 &  12.14 &  34.93 & !7.0  &  $-$0.17 &  !48.3    \\
1558$-$5419  &   6.19 &  12.28 &  33.04 & !9.1  &  $-$0.14 &  !33.1    \\ \\
1601$-$5244  &   7.75 &  12.14 &  30.23 & !5.1  &  +0.01 &  !!3.4    \\
1601$-$5335  &   4.86 &  12.63 &  35.00 & !4.0  &  $-$0.04 &  !!3.6    \\
1605$-$5215  &   6.53 &  12.35 &  32.26 & !7.1  &  +0.00 &  !11.0    \\
1607$-$5140  &   6.33 &  11.97 &  33.40 & !7.0  &  +0.03 &  !12.8    \\
1609$-$5158  &   6.19 &  12.61 &  32.38 & 12.7  &  $-$0.04 &  !43.5    \\ \\
1610$-$5006  &   5.75 &  12.41 &  33.68 & !6.6  &  +0.12 &  !69.5    \\
1611$-$4949  &   7.29 &  11.78 &  31.86 & !8.8  &  +0.18 &  !45.1    \\
1613$-$5211  &   5.58 &  12.48 &  33.90 & !6.2  &  $-$0.08 &  !11.0    \\
1613$-$5234  &   6.19 &  12.32 &  32.97 & !9.9  &  $-$0.19 &  !27.6    \\
1616$-$5109  &   6.01 &  12.69 &  32.62 & 18.9  &  $-$0.11 &  428.7    \\ \\
1616$-$5208  &   5.75 &  12.74 &  33.04 & !7.4  &  $-$0.13 &  !23.9    \\
1621$-$5039  &   6.12 &  12.58 &  32.60 & !4.9  &  $-$0.04 &  !!8.6    \\
1622$-$4802  &   7.14 &  11.46 &  32.81 & !6.0  &  +0.12 &  !33.1    \\
1622$-$4944  &   6.00 &  12.64 &  32.74 & !8.6  &  $-$0.00 &  !38.0    \\
1623$-$4949  &   5.44 &  12.75 &  33.63 & !3.8  &  $-$0.01 &  !!5.1    \\ \\
1625$-$4904  &   5.64 &  12.45 &  33.83 & !7.9  &  +0.02 &  !12.6    \\
1626$-$4807  &   5.43 &  12.36 &  34.43 & 10.2  &  +0.11 &  !38.5    \\
1628$-$4804  &   7.04 &  12.02 &  31.88 & 11.2  &  +0.09 &  125.4    \\
1632$-$4621  &   5.55 &  13.06 &  32.78 & !8.4  &  +0.16 &  !63.8    \\
1632$-$4818  &   4.30 &  13.37 &  34.68 & !8.5  &  $-$0.03 &  !28.4    \\ \\
1649$-$4349  &   8.50 &  11.30 &  30.41 & !5.6  &  +0.06 &  !23.2    \\
1649$-$4729  &   5.86 &  12.15 &  33.99 & 12.7  &  $-$0.39 &  !46.8    \\
1650$-$4502  &   5.57 &  12.40 &  34.04 & !5.1  &  $-$0.03 &  !!9.1    \\
1653$-$4249  &   6.30 &  12.24 &  32.92 & !5.6  &  +0.06 &  !41.4    \\
1709$-$3841  &   6.07 &  12.34 &  33.18 & !5.2  &  +0.08 &  !!8.3    \\ \\
1715$-$3700  &   7.93 &  11.53 &  31.08 & !6.1  &  +0.09 &  !13.6    \\
1716$-$3720  &   5.74 &  12.53 &  33.45 & !9.4  &  +0.09 &  !36.6    \\
1718$-$3825  &   4.95 &  12.00 &  36.11 & !4.2  &  $-$0.03 &  !23.4    \\
1720$-$3659  &   8.23 &  11.03 &  31.48 & !5.1  &  +0.01 &  !19.3    \\
1723$-$3659  &   5.60 &  12.11 &  34.58 & !4.3  &  $-$0.03 &  !27.5    \\ \\
1724$-$3505  &   5.96 &  12.71 &  32.66 & 12.0  &  +0.08 &  !34.6    \\
1725$-$3546  &   6.04 &  12.60 &  32.73 & 10.2  &  $-$0.03 &  !63.5    \\
1726$-$3530  &   4.16 &  13.57 &  34.54 & 10.0  &  $-$0.01 &  !29.9    \\
1726$-$3635  &   6.50 &  11.81 &  33.38 & !7.4  &  $-$0.10 &  !15.8    \\
1728$-$3733  &   8.13 &  11.33 &  31.08 & !4.9  &  $-$0.14 &  !!4.6    \\
\hline
\end{tabular}
\end{minipage}
\end{table*}

\begin{figure} 
\centerline{\psfig{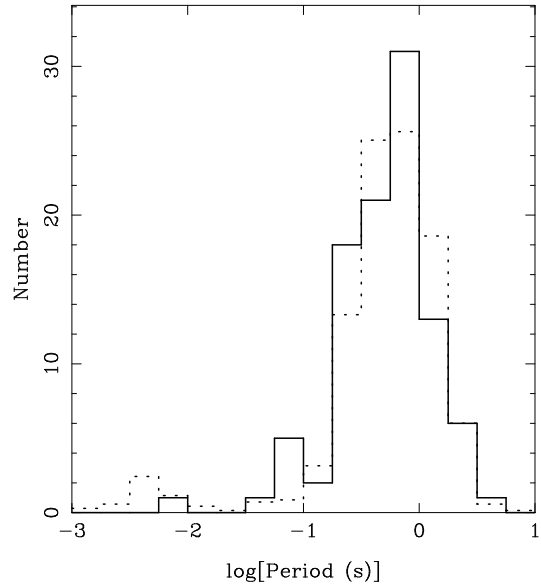}} 
\caption{Distribution in period of the 100 Parkes
multibeam pulsars (solid line) and of previously known pulsars (dotted
line). For the previously known pulsars, the vertical scale has been
adjusted to equalise the number of pulsars in the two distributions. }
\label{fg:phist}
\end{figure}

Fig.~\ref{fg:phist} gives histograms of the distributions in pulse period
for the 100 multibeam pulsars and previously known disk pulsars, i.e.,
excluding those in globular clusters and the Magellanic Clouds. For the
so-called `normal' or non-millisecond pulsars, the distribution of the
multibeam pulsars is similar to that of previously known pulsars, except for
a larger number of pulsars with periods of just less than 100 ms. As shown
by Table~\ref{tb:deriv}, three of these, PSRs J0940$-$5428, J1112$-$6103 and
J1718$-$3825, are relatively young pulsars with ages between 30,000 and
100,000 years and spin-down luminosities in excess of $10^{36}$ erg
s$^{-1}$. The other two, PSRs J1232$-$6501 and J1454$-$5846, have very small
period derivatives and are members of binary systems
(Table~\ref{tb:binpar}). As discussed by Camilo et
al. (2001)\nocite{clm+01}, both of these systems have unusual
properties. The first is atypical of low-mass binary pulsars, having a
relatively long spin period, while the second is unusual in that it has a
larger companion mass and higher eccentricity than most pulsar --
white-dwarf binaries. Eleven of these first 100 pulsars have characteristic
ages of less than 100 kyr; this is a much higher proportion than that for
the previously known population.

Only one millisecond pulsar, PSR J1435$-$6100, which has a period of 9.3 ms
and is a member of a binary system (Table~\ref{tb:binpar}), is included in
first 100 pulsars discovered by the Parkes multibeam survey (although
several more have subsequently been discovered). As Fig.~\ref{fg:phist} shows,
this is a much smaller proportion than that for previously known pulsars,
although it is worth noting that there are no previously known disk
millisecond pulsars within the area currently searched ($|b| \la
1\fdg5$). There are several factors which contribute to this low detection
rate for millisecond pulsars. This paper reports the earliest multibeam
survey observations which were made along and adjacent to the Galactic
equator --- the vast majority of the discovered pulsars have Galactic
latitudes of $\la 1\degr$ (Table~\ref{tb:posn}). At these latitudes, the
volume searched for millisecond pulsars is greatly reduced by dispersion
broadening. Fig.~\ref{fg:sens} shows that the sensitivity is halved for a
10-ms pulsar with DM of 100 cm$^{-3}$ pc, corresponding to a distance of 3
kpc or less in the Galactic plane. The generally lower luminosity of
millisecond pulsars results in a flux-density-limited distribution which
extends to high Galactic latitudes \cite{lml+98}, so the expected number in
our search volume is small. Furthermore, most radio-frequency interference
produces spurious signals at millisecond periods. At the early stage at
which most of these data were processed, techniques for eliminating the
effects of interference were not optimised. Consequently, real pulsars
tended to be lost in a forest of spurious candidates. Finally, many
millisecond pulsars are members of binary systems. The long observation time
of this survey tends to discriminate against detection of short-period
binary systems. All of these factors have been or will be largely overcome
in subsequent observations and analyses.

At the other end of the period range, PSR J1307$-$6318 has a pulse period of
4.96 s, the third longest known. Unlike PSR J2144$-$3933, the 8.5-s
pulsar \cite{ymj99}, PSR J1307$-$6318 has a relatively wide double pulse
(Fig.~\ref{fg:prf}) with a 50 per cent width of 505 ms, more than 10
per cent of the period.

\begin{figure} 
\centerline{\psfig{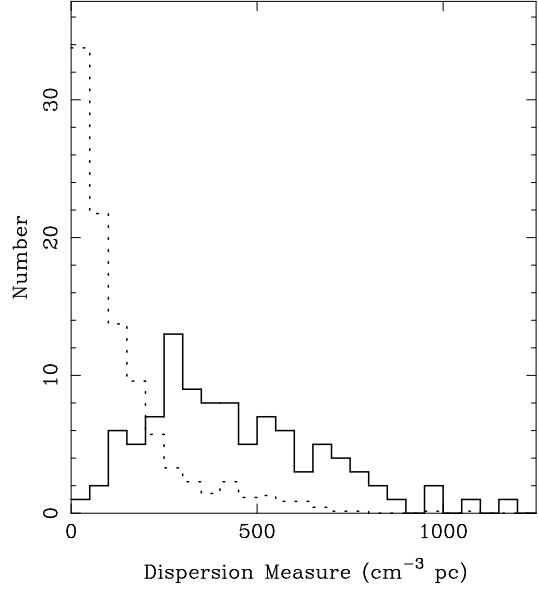}} 
\caption{Distribution in DM of the 100 Parkes
multibeam pulsars (solid line) and of previously known pulsars (dotted
line). For the previously known pulsars, the vertical scale has been
adjusted to equalise the number of pulsars in the two distributions. }
\label{fg:dmhist}
\end{figure}

Fig.~\ref{fg:dmhist} shows that the DM distribution of the multibeam pulsars
is very different from that of previously known pulsars, peaking at a DM of
300 cm$^{-3}$ pc or so. This is readily explained by the low Galactic
latitude and very high sensitivity of the multibeam survey. Most of the
pulsars are distant and of relatively high luminosity
(Table~\ref{tb:deriv}). The Taylor \& Cordes (1993) distance model puts many
of them at distances greater than that of the Galactic Centre, and several
are beyond the limit of the model (those with a distance of 30 kpc in
Table~\ref{tb:deriv}) and certainly over-estimated. Fig.~\ref{fg:prf} shows
that a significant number of these distant pulsars have highly scattered
profiles. However, there is not a close relationship between DM and the
width of the scattering tail, with several pulsars of similar period and
dispersion measure (e.g. PSRs J1609$-$5158 and J1616$-$5109) having quite
different scattering times \cite{man00}.  We expect that the pulsars
discovered in this survey will make a major contribution to improving our
knowledge of the Galactic electron density model and the distribution of the
fluctuations responsible for interstellar scattering, especially in the
central regions of the Galaxy.

\begin{figure} 
\centerline{\psfig{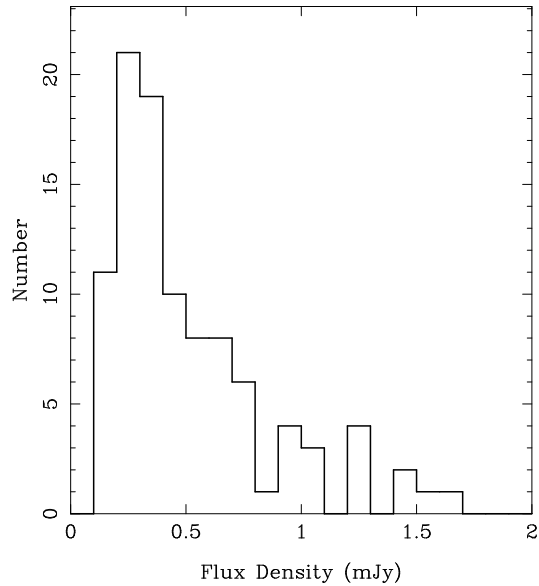}} 
\caption{Distribution in mean flux density at 1400\,MHz of the 100 Parkes
multibeam pulsars.}
\label{fg:s14hist}
\end{figure}

Finally, in Fig.~\ref{fg:s14hist} we show the distribution of mean 1400 MHz
flux densities for the multibeam pulsars. Of the two-thirds of known pulsars
with a published 1400 MHz flux density, only about 10 per cent have a value
of less than 1 mJy. Values above 1 mJy are generally only quoted to the
nearest mJy, so they are not well suited to display in
Fig.~\ref{fg:s14hist}. Ten or so newly discovered
pulsars have $S_{1400} \la 0.2$ mJy, lower
than the nominal survey limiting flux density. Interstellar scintillation is
not normally observed for the pulsars discovered in this survey, as
diffractive scintillation bandwidths are much less than the observed
bandwidth of 288 MHz and refractive scintillations are weak for high-DM
pulsars \cite{ric77,ks92}. The principal reason for the low observed flux
densities is the dependence of effective survey sensitivity on pulse width
(\S\ref{sec:sens}). With only a few exceptions, observed flux densities are
greater than the nominal limiting flux density scaled by
$[W_{50}/(P-W_{50})/0.05]^{1/2}$.  Another factor is that many pulsars show
intrinsic intensity variations such as nulling, and it is likely that some
of these pulsars were detected when they had a greater than average flux
density. As expected, most of the detected pulsars are relatively weak, with
mean flux densities in the range 0.2 to 0.5 mJy. However, because of the
large distances of most of these pulsars, their luminosities are typically
large (Table~\ref{tb:deriv}). All have $L_{1400} > 1$ mJy kpc$^2$,
corresponding to $L_{400} \ga 5$ mJy kpc$^2$ and most are above the
low-luminosity cutoff in the luminosity distribution which, at 400 MHz,
begins at about 10 mJy kpc$^2$ \cite{lml+98}.

The newly discovered pulsars reported in this paper represent only a small
fraction of the total sample which will be discovered by the Parkes
multibeam pulsar survey when it is complete. We therefore defer a more
detailed analysis of the properties of the multibeam sample, its relation to
previously known pulsars and its implications for the Galactic distribution
and evolution of pulsars to later publications.

\section*{Acknowledgements} 
We gratefully acknowledge the technical assistance provided by George Loone,
Tim Ikin, Mike Kesteven, Mark Leach and all of the staff at the Parkes
Observatory toward the development of the Parkes multibeam pulsar system. We
also thank Russell Edwards for providing the program for detecting
narrow-band radio-frequency interference and the Swinburne University of
Technology group led by Matthew Bailes for assistance with development of
the timing analysis software. At various times many people have assisted
with the observing --- we especially thank Paulo Freire, Dominic Morris and
Russell Edwards. FC gratefully acknowledges support from NASA grant
NAG~5-9095 and the European Commission through a Marie Curie fellowship
under contract no. ERB~FMBI~CT961700.  VMK is an Alfred P. Sloan Research
Fellow and was supported in part by a US National Science Foundation Career
Award (AST-9875897) and by a Natural Sciences and Engineering Research
Council of Canada grant (RGPIN 228738-00). IHS received support from NSERC
and Jansky postdoctoral Fellowships. The Parkes radio telescope is
part of the Australia Telescope which is funded by the Commonwealth of
Australia for operation as a National Facility managed by CSIRO.

%\bibliography{journals,modrefs,psrrefs,crossrefs}
%\bibliographystyle{mn}

\end{document}